\documentclass[twocolumn]{aastex62}

\usepackage{epstopdf}
\usepackage{amsmath}
\usepackage{color}
\usepackage{multirow}
\usepackage{gensymb}
\usepackage[normalem]{ulem}
\usepackage{graphicx}
\usepackage{float}

\accepted{April 30, 2019}

\submitjournal{The Astronomical Journal}

\shorttitle{RR Lyrae in DES}
\shortauthors{Stringer, Long, Macri, et al.}

\begin{document}


\title{\sc Identification of RR Lyrae stars in multiband, sparsely-sampled data from the Dark Energy Survey using template fitting and Random Forest classification}

\correspondingauthor{Katelyn M. Stringer}
\email{kstringer@tamu.edu}

\author[0000-0002-4624-2772]{K. M.~Stringer}
\affil{George P.~and Cynthia W.~Mitchell Institute for Fundamental Physics and Astronomy,\\ Department of Physics and Astronomy, Texas A\&M University, College Station, TX 77843, USA}

\author{J. P.~Long}
\affil{Department of Statistics, 3143 TAMU, College Station, TX 77843, USA}

\author{L. M.~Macri}
\affil{George P.~and Cynthia W.~Mitchell Institute for Fundamental Physics and Astronomy,\\ Department of Physics and Astronomy, Texas A\&M University, College Station, TX 77843, USA}

\author{J. L.~Marshall}
\affil{George P.~and Cynthia W.~Mitchell Institute for Fundamental Physics and Astronomy,\\ Department of Physics and Astronomy, Texas A\&M University, College Station, TX 77843, USA}

\author{A. Drlica-Wagner}
\affil{Fermi National Accelerator Laboratory, P. O. Box 500, Batavia, IL 60510, USA}

\author{C. E.~Mart\'inez-V\'azquez}
\affil{Cerro Tololo Inter-American Observatory, National Optical Astronomy Observatory, Casilla 603, La Serena, Chile}

\author{A. K.~Vivas}
\affil{Cerro Tololo Inter-American Observatory, National Optical Astronomy Observatory, Casilla 603, La Serena, Chile}

\author{K. Bechtol}
\affil{Wisconsin IceCube Particle Astrophysics Center (WIPAC), Madison, WI 53703, USA}
\affil{Department of Physics, University of Wisconsin-Madison, Madison, WI 53706, USA}

\author{E. Morganson}
\affil{National Center for Supercomputing Applications, 1205 West Clark St., Urbana, IL 61801, USA}
\affil{Department of Astronomy, University of Illinois, Urbana, IL 61801, USA}

\author{M. Carrasco Kind}
\affil{National Center for Supercomputing Applications, 1205 West Clark St., Urbana, IL 61801, USA}
\affil{Department of Astronomy, University of Illinois, Urbana, IL 61801, USA}

\author{A. B.~Pace}
\affil{George P.~and Cynthia W.~Mitchell Institute for Fundamental Physics and Astronomy,\\ Department of Physics and Astronomy, Texas A\&M University, College Station, TX 77843, USA}

\author{A.~R.~Walker}
\affiliation{Cerro Tololo Inter-American Observatory, National Optical Astronomy Observatory, Casilla 603, La Serena, Chile}

\author{C. Nielsen}
\affil{George P.~and Cynthia W.~Mitchell Institute for Fundamental Physics and Astronomy,\\ Department of Physics and Astronomy, Texas A\&M University, College Station, TX 77843, USA}
\affil{Department of Physics, Purdue University, 1396 Physics Building, West Lafayette, Indiana 47907-1396}

\author{T. S.~Li}
\affil{Fermi National Accelerator Laboratory, P. O. Box 500, Batavia, IL 60510, USA}

\author{E. Rykoff}
\affil{Kavli Institute for Particle Astrophysics \& Cosmology, P. O. Box 2450, Stanford University, Stanford, CA 94305, USA}
\affil{SLAC National Accelerator Laboratory, Menlo Park, CA 94025, USA}

\author{D. Burke}
\affil{Kavli Institute for Particle Astrophysics \& Cosmology, P. O. Box 2450, Stanford University, Stanford, CA 94305, USA}
\affil{SLAC National Accelerator Laboratory, Menlo Park, CA 94025, USA}

\author{A. Carnero Rosell}
\affil{Laborat\'orio Interinstitucional de e-Astronomia - LIneA, Rua Gal. Jos\'e Cristino 77, Rio de Janeiro, RJ - 20921-400, Brazil}
\affil{Observat\'orio Nacional, Rua Gal. Jos\'e Cristino 77, Rio de Janeiro, RJ - 20921-400, Brazil}

\author{E. Neilsen}
\affil{Fermi National Accelerator Laboratory, P. O. Box 500, Batavia, IL 60510, USA}

\author{P. Ferguson}
\affil{George P.~and Cynthia W.~Mitchell Institute for Fundamental Physics and Astronomy,\\ Department of Physics and Astronomy, Texas A\&M University, College Station, TX 77843, USA}

\author{S. A.~Cantu}
\affil{George P.~and Cynthia W.~Mitchell Institute for Fundamental Physics and Astronomy,\\ Department of Physics and Astronomy, Texas A\&M University, College Station, TX 77843, USA}

\author{J. L.~Myron}
\affil{George P.~and Cynthia W.~Mitchell Institute for Fundamental Physics and Astronomy,\\ Department of Physics and Astronomy, Texas A\&M University, College Station, TX 77843, USA}

\author{L. Strigari}
\affil{George P.~and Cynthia W.~Mitchell Institute for Fundamental Physics and Astronomy,\\ Department of Physics and Astronomy, Texas A\&M University, College Station, TX 77843, USA}

\author{A.~Farahi}
\affil{Department of Physics, Carnegie Mellon University, Pittsburgh, Pennsylvania 15312, USA}

\author{F. Paz-Chinch\'on}
\affil{National Center for Supercomputing Applications, 1205 West Clark St., Urbana, IL 61801, USA}
\affil{Department of Astronomy, University of Illinois, Urbana, IL 61801, USA}

\author{D. Tucker}
\affil{Fermi National Accelerator Laboratory, P. O. Box 500, Batavia, IL 60510, USA}

\author{Z. Lin}
\affil{Department of Statistics, 3143 TAMU, College Station, TX 77843, USA}

\author{D. Hatt}
\affil{Kavli Institute for Cosmological Physics, University of Chicago, Chicago, IL 60637, USA}

\author{J. F.~Maner}
\affil{George P.~and Cynthia W.~Mitchell Institute for Fundamental Physics and Astronomy,\\ Department of Physics and Astronomy, Texas A\&M University, College Station, TX 77843, USA}

\author{L. Plybon}
\affil{George P.~and Cynthia W.~Mitchell Institute for Fundamental Physics and Astronomy,\\ Department of Physics and Astronomy, Texas A\&M University, College Station, TX 77843, USA}

\author{A. H.~Riley}
\affil{George P.~and Cynthia W.~Mitchell Institute for Fundamental Physics and Astronomy,\\ Department of Physics and Astronomy, Texas A\&M University, College Station, TX 77843, USA}

\author{E. O.~Nadler}
\affil{Kavli Institute for Cosmological Physics, University of Chicago, Chicago, IL 60637, USA}


\author{T.~M.~C.~Abbott}
\affiliation{Cerro Tololo Inter-American Observatory, National Optical Astronomy Observatory, Casilla 603, La Serena, Chile}

\author{S.~Allam}
\affiliation{Fermi National Accelerator Laboratory, P. O. Box 500, Batavia, IL 60510, USA}

\author{J.~Annis}
\affiliation{Fermi National Accelerator Laboratory, P. O. Box 500, Batavia, IL 60510, USA}

\author{E.~Bertin}
\affiliation{CNRS, UMR 7095, Institut d'Astrophysique de Paris, F-75014, Paris, France}
\affiliation{Sorbonne Universit\'es, UPMC Univ Paris 06, UMR 7095, Institut d'Astrophysique de Paris, F-75014, Paris, France}

\author{D.~Brooks}
\affiliation{Department of Physics \& Astronomy, University College London, Gower Street, London, WC1E 6BT, UK}

\author{E.~Buckley-Geer}
\affiliation{Fermi National Accelerator Laboratory, P. O. Box 500, Batavia, IL 60510, USA}

\author{J.~Carretero}
\affiliation{Institut de F\'{\i}sica d'Altes Energies (IFAE), The Barcelona Institute of Science and Technology, Campus UAB, 08193 Bellaterra (Barcelona) Spain}

\author{C.~E.~Cunha}
\affiliation{Kavli Institute for Particle Astrophysics \& Cosmology, P. O. Box 2450, Stanford University, Stanford, CA 94305, USA}

\author{C.~B.~D'Andrea}
\affiliation{Department of Physics and Astronomy, University of Pennsylvania, Philadelphia, PA 19104, USA}

\author{L.~N.~da Costa}
\affiliation{Laborat\'orio Interinstitucional de e-Astronomia - LIneA, Rua Gal. Jos\'e Cristino 77, Rio de Janeiro, RJ - 20921-400, Brazil}
\affiliation{Observat\'orio Nacional, Rua Gal. Jos\'e Cristino 77, Rio de Janeiro, RJ - 20921-400, Brazil}

\author{J.~De~Vicente}
\affiliation{Centro de Investigaciones Energ\'eticas, Medioambientales y Tecnol\'ogicas (CIEMAT), Madrid, Spain}

\author{S.~Desai}
\affiliation{Department of Physics, IIT Hyderabad, Kandi, Telangana 502285, India}

\author{P.~Doel}
\affiliation{Department of Physics \& Astronomy, University College London, Gower Street, London, WC1E 6BT, UK}

\author{T.~F.~Eifler}
\affiliation{Department of Astronomy/Steward Observatory, 933 North Cherry Avenue, Tucson, AZ 85721-0065, USA}
\affiliation{Jet Propulsion Laboratory, California Institute of Technology, 4800 Oak Grove Dr., Pasadena, CA 91109, USA}

\author{B.~Flaugher}
\affiliation{Fermi National Accelerator Laboratory, P. O. Box 500, Batavia, IL 60510, USA}

\author{J.~Frieman}
\affiliation{Fermi National Accelerator Laboratory, P. O. Box 500, Batavia, IL 60510, USA}
\affiliation{Kavli Institute for Cosmological Physics, University of Chicago, Chicago, IL 60637, USA}

\author{J.~Garc\'ia-Bellido}
\affiliation{Instituto de Fisica Teorica UAM/CSIC, Universidad Autonoma de Madrid, 28049 Madrid, Spain}

\author{E.~Gaztanaga}
\affiliation{Institut d'Estudis Espacials de Catalunya (IEEC), 08034 Barcelona, Spain}
\affiliation{Institute of Space Sciences (ICE, CSIC),  Campus UAB, Carrer de Can Magrans, s/n,  08193 Barcelona, Spain}

\author{D.~Gruen}
\affiliation{Department of Physics, Stanford University, 382 Via Pueblo Mall, Stanford, CA 94305, USA}
\affiliation{Kavli Institute for Particle Astrophysics \& Cosmology, P. O. Box 2450, Stanford University, Stanford, CA 94305, USA}
\affiliation{SLAC National Accelerator Laboratory, Menlo Park, CA 94025, USA}

\author{J.~Gschwend}
\affiliation{Laborat\'orio Interinstitucional de e-Astronomia - LIneA, Rua Gal. Jos\'e Cristino 77, Rio de Janeiro, RJ - 20921-400, Brazil}
\affiliation{Observat\'orio Nacional, Rua Gal. Jos\'e Cristino 77, Rio de Janeiro, RJ - 20921-400, Brazil}

\author{G.~Gutierrez}
\affiliation{Fermi National Accelerator Laboratory, P. O. Box 500, Batavia, IL 60510, USA}

\author{W.~G.~Hartley}
\affiliation{Department of Physics \& Astronomy, University College London, Gower Street, London, WC1E 6BT, UK}
\affiliation{Department of Physics, ETH Zurich, Wolfgang-Pauli-Strasse 16, CH-8093 Zurich, Switzerland}

\author{D.~L.~Hollowood}
\affiliation{Santa Cruz Institute for Particle Physics, Santa Cruz, CA 95064, USA}

\author{B.~Hoyle}
\affiliation{Max Planck Institute for Extraterrestrial Physics, Giessenbachstrasse, 85748 Garching, Germany}
\affiliation{Universit\"ats-Sternwarte, Fakult\"at f\"ur Physik, Ludwig-Maximilians Universit\"at M\"unchen, Scheinerstr. 1, 81679 M\"unchen, Germany}

\author{D.~J.~James}
\affiliation{Harvard-Smithsonian Center for Astrophysics, Cambridge, MA 02138, USA}

\author{K.~Kuehn}
\affiliation{Australian Astronomical Optics, Macquarie University, North Ryde, NSW 2113, Australia}

\author{N.~Kuropatkin}
\affiliation{Fermi National Accelerator Laboratory, P. O. Box 500, Batavia, IL 60510, USA}

\author{P.~Melchior}
\affiliation{Department of Astrophysical Sciences, Princeton University, Peyton Hall, Princeton, NJ 08544, USA}

\author{R.~Miquel}
\affiliation{Instituci\'o Catalana de Recerca i Estudis Avan\c{c}ats, E-08010 Barcelona, Spain}
\affiliation{Institut de F\'{\i}sica d'Altes Energies (IFAE), The Barcelona Institute of Science and Technology, Campus UAB, 08193 Bellaterra (Barcelona) Spain}

\author{R.~L.~C.~Ogando}
\affiliation{Laborat\'orio Interinstitucional de e-Astronomia - LIneA, Rua Gal. Jos\'e Cristino 77, Rio de Janeiro, RJ - 20921-400, Brazil}
\affiliation{Observat\'orio Nacional, Rua Gal. Jos\'e Cristino 77, Rio de Janeiro, RJ - 20921-400, Brazil}

\author{A.~A.~Plazas}
\affiliation{Jet Propulsion Laboratory, California Institute of Technology, 4800 Oak Grove Dr., Pasadena, CA 91109, USA}

\author{E.~Sanchez}
\affiliation{Centro de Investigaciones Energ\'eticas, Medioambientales y Tecnol\'ogicas (CIEMAT), Madrid, Spain}

\author{B.~Santiago}
\affiliation{Instituto de F\'\i sica, UFRGS, Caixa Postal 15051, Porto Alegre, RS - 91501-970, Brazil}
\affiliation{Laborat\'orio Interinstitucional de e-Astronomia - LIneA, Rua Gal. Jos\'e Cristino 77, Rio de Janeiro, RJ - 20921-400, Brazil}

\author{V.~Scarpine}
\affiliation{Fermi National Accelerator Laboratory, P. O. Box 500, Batavia, IL 60510, USA}

\author{M.~Schubnell}
\affiliation{Department of Physics, University of Michigan, Ann Arbor, MI 48109, USA}

\author{S.~Serrano}
\affiliation{Institut d'Estudis Espacials de Catalunya (IEEC), 08034 Barcelona, Spain}
\affiliation{Institute of Space Sciences (ICE, CSIC),  Campus UAB, Carrer de Can Magrans, s/n,  08193 Barcelona, Spain}

\author{I.~Sevilla-Noarbe}
\affiliation{Centro de Investigaciones Energ\'eticas, Medioambientales y Tecnol\'ogicas (CIEMAT), Madrid, Spain}

\author{M.~Smith}
\affiliation{School of Physics and Astronomy, University of Southampton,  Southampton, SO17 1BJ, UK}

\author{R.~C.~Smith}
\affiliation{Cerro Tololo Inter-American Observatory, National Optical Astronomy Observatory, Casilla 603, La Serena, Chile}

\author{M.~Soares-Santos}
\affiliation{Brandeis University, Physics Department, 415 South Street, Waltham MA 02453}

\author{F.~Sobreira}
\affiliation{Instituto de F\'isica Gleb Wataghin, Universidade Estadual de Campinas, 13083-859, Campinas, SP, Brazil}
\affiliation{Laborat\'orio Interinstitucional de e-Astronomia - LIneA, Rua Gal. Jos\'e Cristino 77, Rio de Janeiro, RJ - 20921-400, Brazil}

\author{E.~Suchyta}
\affiliation{Computer Science and Mathematics Division, Oak Ridge National Laboratory, Oak Ridge, TN 37831}

\author{M.~E.~C.~Swanson}
\affiliation{National Center for Supercomputing Applications, 1205 West Clark St., Urbana, IL 61801, USA}

\author{G.~Tarle}
\affiliation{Department of Physics, University of Michigan, Ann Arbor, MI 48109, USA}

\author{D.~Thomas}
\affiliation{Institute of Cosmology and Gravitation, University of Portsmouth, Portsmouth, PO1 3FX, UK}

\author{V.~Vikram}
\affiliation{Argonne National Laboratory, 9700 South Cass Avenue, Lemont, IL 60439, USA}

\author{B.~Yanny}
\affiliation{Fermi National Accelerator Laboratory, P. O. Box 500, Batavia, IL 60510, USA}

\collaboration{(DES Collaboration)}


\begin{abstract}

Many studies have shown that RR Lyrae variable stars (RRL) are powerful stellar tracers of Galactic halo structure and satellite galaxies. The Dark Energy Survey (DES), with its deep and wide coverage ($g\sim 23.5$~mag in a single exposure; over 5000 deg$^{2}$) provides a rich opportunity to search for substructures out to the edge of the Milky Way halo. However, the sparse and unevenly sampled multiband light curves from the DES wide-field survey (median 4 observations in each of \textit{grizY} over the first three years) pose a challenge for traditional techniques used to detect RRL. We present an empirically motivated and computationally efficient template fitting method to identify these variable stars using three years of DES data.  When tested on DES light curves of previously classified objects in SDSS stripe 82, our algorithm recovers 89\% of RRL periods to within 1\% of their true value with 85\% purity and 76\% completeness. Using this method, we identify 5783 RRL candidates, $\sim31\%$ of which are previously undiscovered. This method will be useful for identifying RRL in other sparse multiband data sets.  

\end{abstract}

\keywords{stars: variable stars: RR Lyrae --- Galaxy: structure --- Galaxy: halo}

\reportnum{DES-2018-0375}
\reportnum{FERMILAB-PUB-18-680-AE}


\vfill
\section{Introduction} \label{sec:intro}

RR Lyrae variable stars (RRL) are old (age $>10$ Gyr) horizontal branch stars that pulsate with short periods (0.2 - 1.2 days). They have become one of the most widely used stellar tracers in Milky Way and Local Group studies. Thanks to the discovery of RR Lyrae itself \citep{pickering1901} and the subsequent studies of their pulsation (see \citealt{kingcox1968} and \citealt{catelansmith2015} for a review of the pioneering and current works in this field), these stars have well-understood period-luminosity-metallicity (\textit{P-L-Z}) relations\footnote{These are sometimes presented as Period-Luminosity-Color \textit{(P-L-C)} relations.} \citep[e.g.,][]{caceres2008,marconi2015}, making them excellent distance indicators, especially in the near-infrared bands. This, combined with their bright luminosities ($M_{V}\sim 0.6$) and advanced ages make RRL well-suited to trace discrete stellar populations (satellite galaxies, star clusters, and streams) within the Milky Way halo \citep[e.g.,][]{catelan2004,vivas2004,caceres2008,sesar2010,stetson2014,fiorentino2015}.  

Locating these stellar populations is crucial for testing the $\Lambda$CDM hierarchical model, which predicts that the haloes of large galaxies like the Milky Way are formed through the accretion and disruption of lower mass haloes \citep{bullock2005lcdm}. Recent re-examinations of these simulations predict that the outer reaches of the stellar halo ($d\ge 100$~kpc) are primarily composed of the most recently accreted satellites and that thousands of RRL should be present in them \citep{sanderson2017}. Once satellite galaxies and their disrupted remains are found, their distribution and properties can reveal valuable clues about the formation history, dark matter density profile, and mass of the Milky Way. While these objects are interesting in their own right, the statistical information about this sample is vital to place the Milky Way in a broader cosmological context.

Numerous Milky Way substructures have already been discovered. Eleven ``classical'' dwarf galaxies were known to orbit the Milky Way before 2005 \citep{mcconnachie2012}\footnote{The nature of the Canis Major Overdensity as a satellite galaxy is in doubt due to a lack of an RRL excess and a potential warp in the Milky Way disk \citep{mateu2009}.}. Thanks to the advent of wide-field surveys such as the Sloan Digital Sky Survey \citep[SDSS,][]{york2000sdss}, the Panoramic Survey Telescope and Rapid Response System \citep[Pan-STARRS,][]{chambers2016panstarrs}, and the Dark Energy Survey \citep[DES,][]{des2016}, over 40 new dwarf satellite candidates have been discovered \citep[][]{willman2005dwarfa,willman2005dwarfb,zucker2006dwarfa,zucker2006dwarfb,belokurov2006dwarf,belokurov2007dwarf,belokurov2008dwarf,belokurov2009dwarf,belokurov2010dwarf,grillmair2006dwarf,grillmair2009dwarf,sakamoto2006dwarf,irwin2007dwarf,walsh2007dwarf,kim2015,bechtol2015,koposov2015,drlica-wagner2015,laevens2015a,laevens2015b,luque2016des1,torrealba2016a,luque2017dwarf,torrealba2016b,koposov2018,torrealba2018antlia}. In addition to galaxies that are still intact, tidal streams, the disrupted remains of satellite galaxies and globular clusters, have been discovered to be prevalent within the Milky Way halo \citep[e.g.,][]{odenkirchen2001,newberg2002,belokurov2006,bernard2014,koposov2014,grillmair2016,bernard2016,balbinot2016,shipp2018,mateu2018galstreams}. Besides these, additional large stellar overdensities populate the MW stellar halo with origins still unknown \citep[e.g.,][]{vivas2001,newberg2002,majewski2004,rochapinto2004,sesar2007,belokurov2007overdensity,sharma2010,deason2014,li2016,pieres2017,bergemann2018,prudil2018}.

Most of these satellites and streams were discovered as stellar overdensities in photometric catalogs \citep[][and references therein]{willman2010}. However, this detection method is biased against diffuse objects with low surface brightness \citep[$\mu_{V,0} \gtrsim 29\  \mathrm{mag / arcsec}^{2}$;][]{baker2015}, so an alternative method is needed to locate other faint structures that may have evaded detection. RRL are sufficiently rare so as to not randomly form in pairs outside of stellar structures, so searching for groups of spatially close RRL provides an independent method to detect new structures \citep{ivezic2004trace,sesar2014,baker2015,medina2017,medina2018}. Indeed, at least one RRL has been found in almost every satellite galaxy with available time series data\footnote{One notable exception is the satellite galaxy candidate Carina III, which currently has no detected RRL in its vicinity \citep{torrealba2018}.} \citep[][and references therein]{boettcher2013,vivas2016,martinezvazquez2017}. Thus, identifying RRL in the halo can increase the census of old, metal poor satellite galaxies, streams, and overdensities, and improve our understanding of the Milky Way. 

The two most common subtypes of RRL are those pulsating in the fundamental mode, RRab, and those pulsating in the first overtone, RRc. When their light curves are adequately sampled, RRab are easily identified by their short periods ($0.4 \lesssim P \lesssim 1$~d), relatively large pulsation amplitudes ($0.5 \lesssim A_g \lesssim 1.5$~mag), and a characteristic sawtooth shape. RRc have shorter periods ($0.2 \lesssim P \lesssim0.45$~d), smaller amplitudes ($0.2 \lesssim A_g \lesssim 0.8$~mag), more sinusoidal-shaped light curves, and are generally less numerous than RRab. The fraction of RRab to RRc and the average periods of each are highly dependent on the metallicity of the stellar population in which they formed and is still not fully understood (see \citealt{catelan2009} and references therein.) Most populations of RRL in the Milky Way are commonly subdivided into Oosterhoff I, II, and III groups based on these observational properties (named after the first dichotomy applied to globular clusters by \citealt{oosterhoff1939}. We refer the interested reader to Table 6 in \citet{martinezvazquez2017} for a summary of these properties for a selection of Local Group dwarf galaxies. 

Period-finding algorithms have long been used in conjunction with visual inspection to identify RRL from their time series photometry. However, with the dramatic increases in available data in recent years, the need for automated detection algorithms has grown significantly. \citet{stetson1996} made great strides in this regard when he introduced an automated method to identify Cepheid variables using template light curves to estimate their periods and a scoring system based on calculated variability indices. Recent studies have extended this period-finding technique to multiple filters \citep{mateu2012,vanderplas2015,mondrik2015,saha2017}. However, even these algorithms suffer in performance when applied to extremely sparsely-sampled data. \citet{hernitschek2016} and \citet{sesar2017} developed separate techniques to identify RRL in the sparsely-sampled multiband Pan-STARRS data \citep{chambers2016panstarrs} and found thousands of such variables.

We add to this census by presenting new RRL candidates discovered in the first three years of the DES data. DES is a five-year multiband (\textit{grizY}) imaging survey using the Dark Energy Camera \citep{decam2015} on the 4-m Blanco Telescope at Cerro Tololo Inter-American Observatory (CTIO). After the conclusion of its observations, DES will provide a deep ($\sim25$~mag in the final coadded images) and wide ($\sim 5000\deg^2$) dataset near the Southern Galactic cap \citep{des2005des,diehl2016des}. By the end of the survey, the entire footprint will have been imaged $\sim10$ times in each band. While the main goal of the survey is to better constrain certain cosmological parameters, the deep and wide survey data provide an excellent test bed for probing Milky Way substructure with RRL. However, like Pan-STARRS, the DES light curves are multiband and poorly sampled. In this paper, we detail how we overcome these challenges by creating a empirically derived light curve template and a computationally efficient fitting algorithm to determine periods and other light curve parameters. We use these methods to identify $5783$ RRL candidates, 31\% of which are new discoveries, including three with a heliocentric distance $>$220 kpc. This novel technique will prove useful for other sparsely sampled multiband data sets. 

The rest of the paper is organized as follows: \S\ref{sec:data} describes how we extracted star-like objects from the DES Y3 data release; \S\ref{sec:var} explains how we rescaled the photometric uncertainties and applied simple metrics to select variable objects; \S\ref{sec:template} presents the multiband RRL template, its application to DES light curves, and the construction and performance of our random forest classifier; \S\ref{sec:catalog} presents our RRL catalog, a comparison to overlapping surveys, and parameter uncertainties; \S\ref{sec:discussion} discusses possible biases, the spatial distribution of the candidates, and potential future application for LSST.


\section{Data} \label{sec:data}
\subsection{DES Year 3 Quick Release}\label{sec:y3q2}

\begin{figure*}[htb]
  \includegraphics[width=1\textwidth]{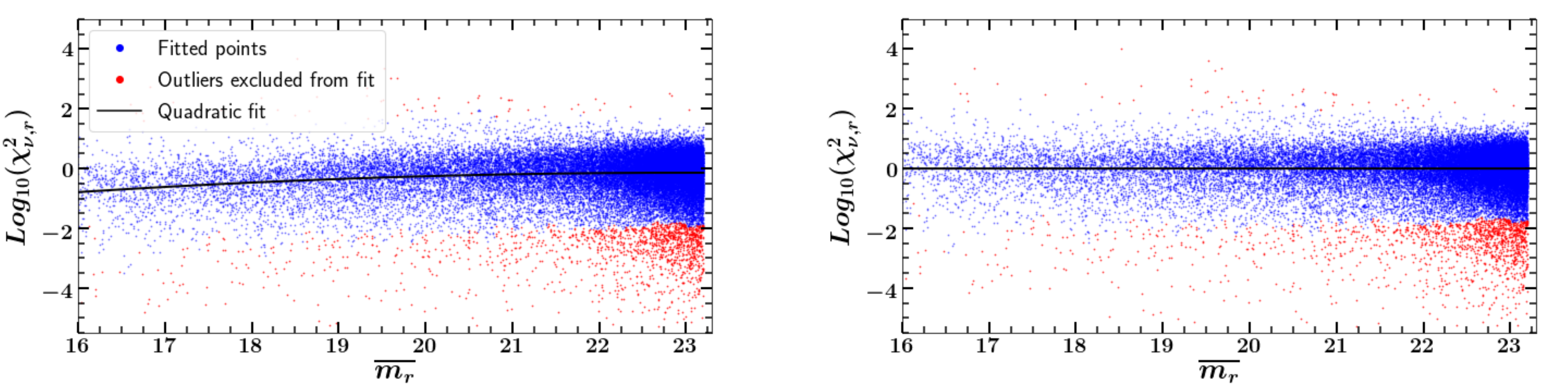}
  \caption{\textit{Left:} Variation of $\log_{10}(\chi_{\nu,r}^{2})$ vs.~median $r$ magnitude, $\overline{m_{r}}$, demonstrating that photometric errors are slightly overestimated for brighter objects in the DES pipeline. Red points were excluded using an iterative 3-$\sigma$ clipping procedure. The black curve shows the quadratic fit that was used to rescale the errors. \textit{Right:} Distribution after the photometric errors were rescaled.}
  \label{fig:rescale}
\end{figure*}

This work is based on the DES internal Year 3 Quick Release catalog (hereafter Y3Q2), which contains all the single epoch data from years 1-3 that formed the basis for the coadded DES first public data release\footnote{https://des.ncsa.illinois.edu/releases/dr1} \citep[][hereafter DR1]{abbott2018}. The Y3Q2 data set was developed in the same manner as the Y2Q1 data release used for the stellar overdensity searches in \citet{drlica-wagner2015} (see their \S2.1 for a detailed description of how the Quick Release catalogs were generated), with one major change. Instead of using stellar locus regression \citep[SLR][]{ivezic2004,macdonald2004,high2009,gilbank2011,desai2012,coupon2012,kelly2014} to determine zeropoints for the absolute photometric calibration, the Y3Q2 release utilizes the Forward Global Calibration Module (FGCM) photometric zeropoints \citep{burke2018}. Y3Q2 contains single epoch catalogs generated from the reduced FINALCUT DES images \citep{morganson2018} and a cross-matched ``coadded catalog" generated from these single epoch measurements. This catalog does not contain information from exposures in which an object was not detected at approximately a 5-$\sigma$ level. The DES Y3Q2 coadded catalog contains nearly $2.9\times 10^8$ unique objects and spans the entire survey footprint with $S/N \sim10$ at a median depth of 23.5, 23.3, 22.8, 22.1, and 20.7~mag in \textit{grizY}, respectively \citep{abbott2018}.

As DES images are collected, the filter to be used and the location to be imaged are prioritized according to the time of the year, the sky conditions (Moon phase, seeing, weather), and how many times that particular area has already been imaged \citep{neilsen2014obstac} \citep[see Fig.~3 in][]{diehl2016des}. While this strategy ensures uniform depth and the best use of the observing time, objects in the wide-field survey are sampled with a highly unpredictable cadence. In the Y3Q2 data set, individual objects can have from 2 to over 50 observations depending on their location. We ensure that each light curve only contains photometric observations by requiring that each observation has a SExtractor warning value $\mathrm{FLAG} \le4$, is sufficiently far away from masked regions in the images ($\mathrm{IMAFLAG\_ISO} \le4$), and has a zeropoint correction available ($\mathrm{FGCM\_FLAG}\le4$). After these cuts, the median number of total observations for a given object is 10, while the median number of observations in each band across the survey region is 4. The effects of the survey coverage and these cuts are discussed more in \S\ref{sec:detection_biases}.

\subsection{Object Selection}\label{sec:selection}

We selected our objects using the coadded catalog before examining the time series data, since the former contained most of the information needed to identify candidates (such as the number of times each object was imaged in each band and the star-galaxy classification). We further restricted the sample to stellar-like objects by following a prescription similar to \citet{bechtol2015}, based on the SExtractor \citep{sextractor1996} {\sc spread\_model} parameter which selects stars well down to $r\sim23$ \citep{drlica-wagner2015}. Lastly, we required at least five total observations to be able to search for variability.

We selected objects that are bright enough to be detected in multiple images by requiring the coadded PSF (\textsc{WAVG\_MAG\_PSF}) or the aperture magnitudes in the exposure with the best seeing in that band (\textsc{MAG\_AUTO}) to be brighter than the median depth of the Y3Q2 single-epoch exposures across the entire survey region (see \S\ref{sec:y3q2}). We excluded all objects for which the coadded photometry errors exceed 0.3 mag in each of \textit{griz} to reduce the number of spurious detections.

To ensure that we did not discard stars with missing data in a single band, we considered these quantities separately for each of \textit{griz}. We did not use the \textit{Y} data for these initial selections because those exposures are generally taken under worse seeing conditions than the other bands \citep{diehl2016des} and are thus a poor choice to use for star-galaxy separation. The star-galaxy separation we used performs best in $riz$ due to the better seeing conditions for those observations, as discussed in the previous section. Including objects which passed this cut in $g$ likely allowed some extended sources into our sample, which we discuss further in \S\ref{sec:visual_validation}. 

Although RRL have well-characterized colors, we did not employ a color cut in this early stage of the analysis because we did not want to exclude any potential RRL with poor coverage across filters or pulsation phase. Simultaneous colors were not available for some objects in the DES footprint, so we would have to calculate colors using coadded magnitudes or magnitudes from arbitrary phases in the star's variation to calculate colors, which would expand the range of possible colors for RRL in our data. The RRL template we describe in \S\ref{sec:template_description} provides the color information we need to identify RRL candidates. In the future, when more epochs of DES data are available, color cuts will be a more reliable RRL indicator prior to the template fitting.
 
In summary, our combined selection criteria were:
\begin{itemize}
    \setlength\itemsep{0.2em}
	\item{$\geq$ 2 observations in \textit{g,r,i,} or \textit{z}}
	\item $|\textsc{spread\_model}| \leq (0.003 + \\
    $\textsc{spreaderr\_model}) in \textit{g,r,i,} or \textit{z}
	\item $ (16 \leq \textsc{wavg\_mag\_psf} \leq \textrm{median depth}) $ or \\
    $ (16 \leq \textsc{mag\_auto} \leq \textrm{median depth}) $ in \textit{g,r,i,} or \textit{z}
	\item $ \textsc{(wavg\_magerr\_psf} < 0.3 )$ or \\
    $ (\textsc{magerr\_auto} < 0.3 )$ in \textit{g,r,i,} or \textit{z}
\end{itemize}

A sample of $\sim 1.5\times 10^8$ objects passed all of these combined selection criteria. We used their time series data instead of their coadded values for the remainder of our analysis. 


\section{Variability Analysis} \label{sec:var}

\subsection{Error Rescaling}
\label{sec:rescale}

Photometric uncertainties can have a large impact on the success of our variability classification. We first account for both the photometric uncertainties reported by the DES pipeline and the uncertainties in the FGCM zeropoint solution\footnote{At the time of this analysis, only a pre-release version of the FGCM zeropoints was available. A later version of these zeropoints was used for other DES Year 3 analyses.} for each exposure by adding them in quadrature. 
Since photometric uncertainties can be over- or under-estimated for different magnitude ranges \citep{kaluzny1998}, we calculated the reduced chi-squared statistic, $\chi_{\nu,b}^{2}$, from the median magnitude $\overline{m_{b}}$ in a given band $b$ for each light curve:     
\begin{equation}
   \chi_{\nu,b}^{2} = \frac{1}{N_{b}-1}\sum_{1}^{N_{b}}\frac{(m_{i,b}-\overline{m_{b}})^{2}}{\sigma_{i,b}^{2}}.
   \end{equation}

\noindent where $N_{b}$ is the number of observations for a unique object in band \textit{b}, ${m_{i,b}}$ is the $i^{th}$ observation in that band, and $\sigma_{i,b}$ is the photometric uncertainty combined with the zeropoint uncertainty for that observation. As this statistic measures the goodness-of-fit to a constant value of $\overline{m_{b}}$, one would expect $\chi_{\nu,b}^{2}\approx 1$ for a non-variable source and $\chi_{\nu,b}^{2}> 1$ for variable sources.

\begin{figure}[tb]
  \includegraphics[width=0.5\textwidth]{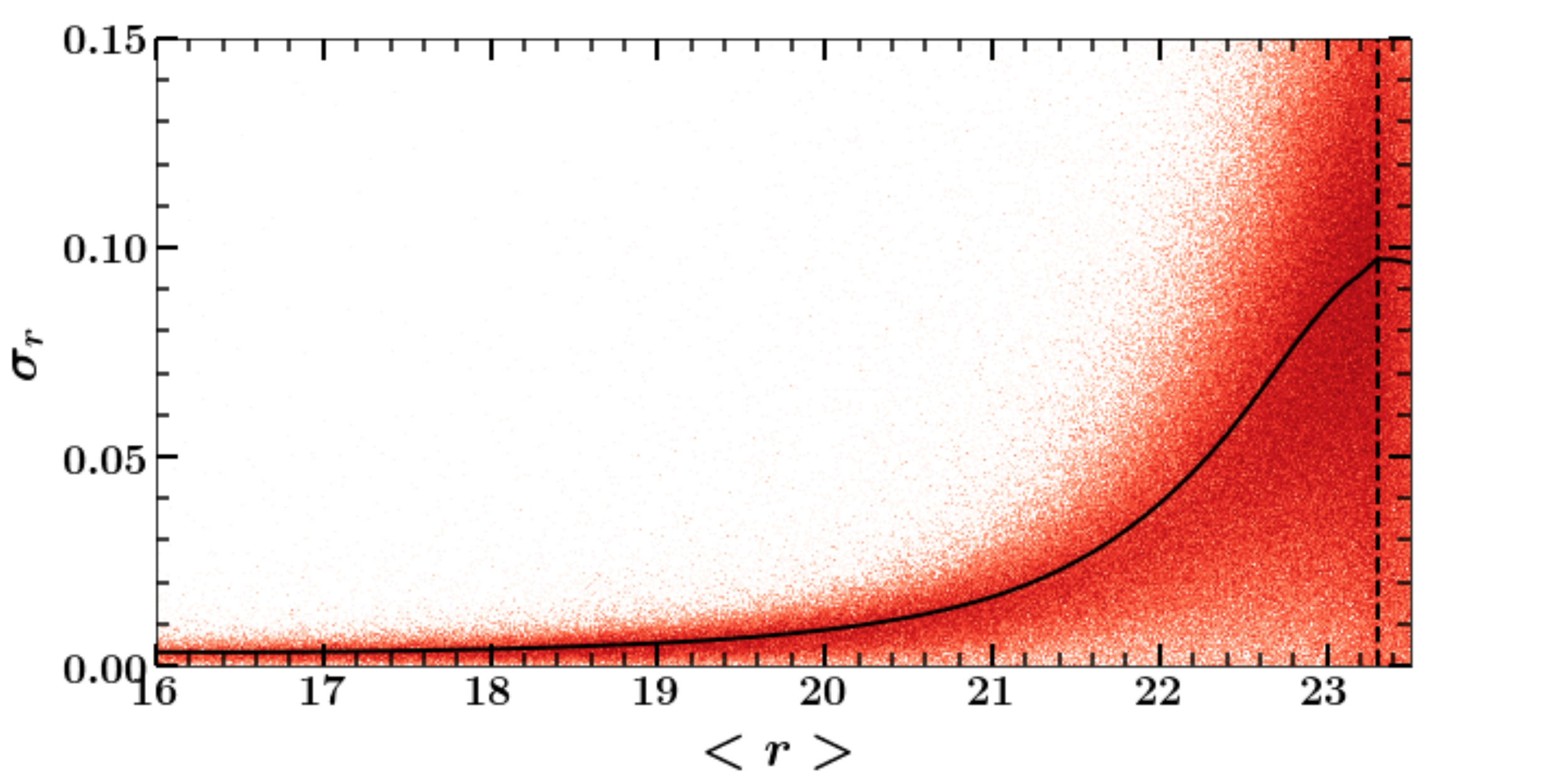}
  \caption{Uncertainties in the $r$-band magnitudes for the entire survey region after the error rescaling described in \S\ref{sec:rescale}. The dashed black line denotes the median survey depth in this band and the solid black line shows the 3rd degree polynomial fit to the uncertainties.}
  \label{fig:rescale_errors}
\end{figure}

Since the majority of objects within a given field will have constant (non-varying) light curves, any overall trend of $\chi_{\nu,b}^{2}$ vs.~magnitude will be indicative of incorrect estimations of photometric uncertainty. For ease of calculations, we subdivided the single epoch data by HEALPix (nside=32) \citep{gorski2005}\footnote{These HEALPix indices are also provided in the DES DR1 products \citep{abbott2018}.} and filter. This resulted in 1772 unique DES HEALPix regions. We fit a quadratic function: 
\begin{equation}
\mathrm{log}_{10}(\chi_{\nu,b}^{2}) = c_{0,b} + c_{1,b}(\overline{m_{b}}-20) + c_{2,b}(\overline{m_{b}}-20)^{2} 
\label{eq:quadfit}
\end{equation}

\noindent for each filter $b$ in each of these regions, excluding variables and outliers by applying an iterative 3-$\sigma$ clip from the median value using the {\tt sigma\_clip} function in {\tt astropy.stats}.  We show the initial trend in $\chi_{\nu,r}^{2}$ for the objects in HEALPix 11678 in the \textit{r} band in the left portion of Figure \ref{fig:rescale}. We multiplied the reported photometric uncertainties of each object by scaling factors based on the best-fit value of $\chi_{\nu,b}^{2}$ for each of its magnitudes in a given band. Once the uncertainties were suitably rescaled, we repeated the calculation to verify that no trends remained. The right portion of Figure \ref{fig:rescale} shows the resulting lack of trend in $\chi_{\nu,r}^{2}$ after the rescaling procedure. Table \ref{tab:chi2coeffs} lists the coefficients used to rescale the errors in each band for this example region (the full version is available online). Figure \ref{fig:rescale_errors} shows the final rescaled photometric uncertainties for the entire survey region as a function of magnitude for the \textit{r} band.
\begin{deluxetable}{lcllr}
	\tabletypesize{\footnotesize}
    
	\tablecaption{Error rescaling coefficients\\
    for Equation \ref{eq:quadfit}}
    \tablehead{
    \colhead{HEALPix} & \colhead{Band $b$} & \colhead{$c_{0,b}$}&\colhead{$c_{1,b}$}&\colhead{$c_{2,b}$}}
	\startdata
          & $g$ & -0.2672 & 0.0816 & -0.0152 \\
          & $r$ & -0.2572 & 0.0793 & -0.0134 \\
    11678 & $i$ & -0.1180 & 0.0395 & -0.0156 \\
          & $z$ & -0.2160 & 0.0827 & -0.0198 \\
          & $Y$ & -0.1942 & 0.0505 & -0.0250 \\
	\enddata
    \tablecomments{The full version of this table is available online.}
    \label{tab:chi2coeffs}
\end{deluxetable}

\subsection{Variability Cuts}
\label{sec:varcuts}

Once photometric uncertainties were rescaled, we assessed the variability of each light curve using two simple metrics. The first was $\chi_{\nu,b}^{2}$ described in \S\ref{sec:rescale}. The second was a metric we called ``significance'', consisting of a weighted range of the magnitudes in one band that acts as a proxy for light curve amplitude\footnote{While other metrics such as the Welch-Stetson I \cite{welch-stetson1993} and Stetson J \citep{stetson1996} indices are widely used and very effective at detecting variability, due to the sparsity of our observations, we chose to use metrics that were agnostic of observation time.}.  

\begin{equation}
   \label{eq:sig}
   \textrm{significance}_{b} = \frac{(m_{\textrm{max},b}-m_{\textrm{min},b})}{\sqrt{\sigma_{\textrm{max},b}^{2}+\sigma_{\textrm{min},b}^{2}}}
   \end{equation}

To test the effectiveness of these metrics and determine the threshold values to separate variables from constant stars, we assembled a labeled training set of previously classified objects in SDSS stripe 82 region (hereafter, ``S82''). S82 is a $\sim 300 \deg^{2}$ area spanning $300^{\circ} \lesssim \alpha \lesssim 60^{\circ}$ and $|\delta|\lesssim 1.25^{\circ}$ that was observed 70-90 times by SDSS in \textit{ugriz} over a period of 10 years. Numerous authors used the resultant well-sampled multiband light curves to identify thousands of variable stars in the region with high confidence \citep{ivezic2007,sesar2010,suveges2012}. These labeled objects are extremely useful for studies of variables from both hemispheres thanks to their equatorial location. Although the magnitude range of DES is deeper than that of SDSS, there is sufficient overlap to create a well-populated training set for our study. Using the calibration and variable catalogs from \citet{ivezic2007}, we cross-matched 641,710 ``standard'' (i.e., constant) stars and 16,752 variables in common between SDSS and DES objects. We also identified 296 RRL in common between DES and either \citet{sesar2010} or \citet{suveges2012}, consisting of 238 and 58 objects of subtype RRab and RRc, respectively.

\begin{figure}[]
  \includegraphics[width=0.5\textwidth]{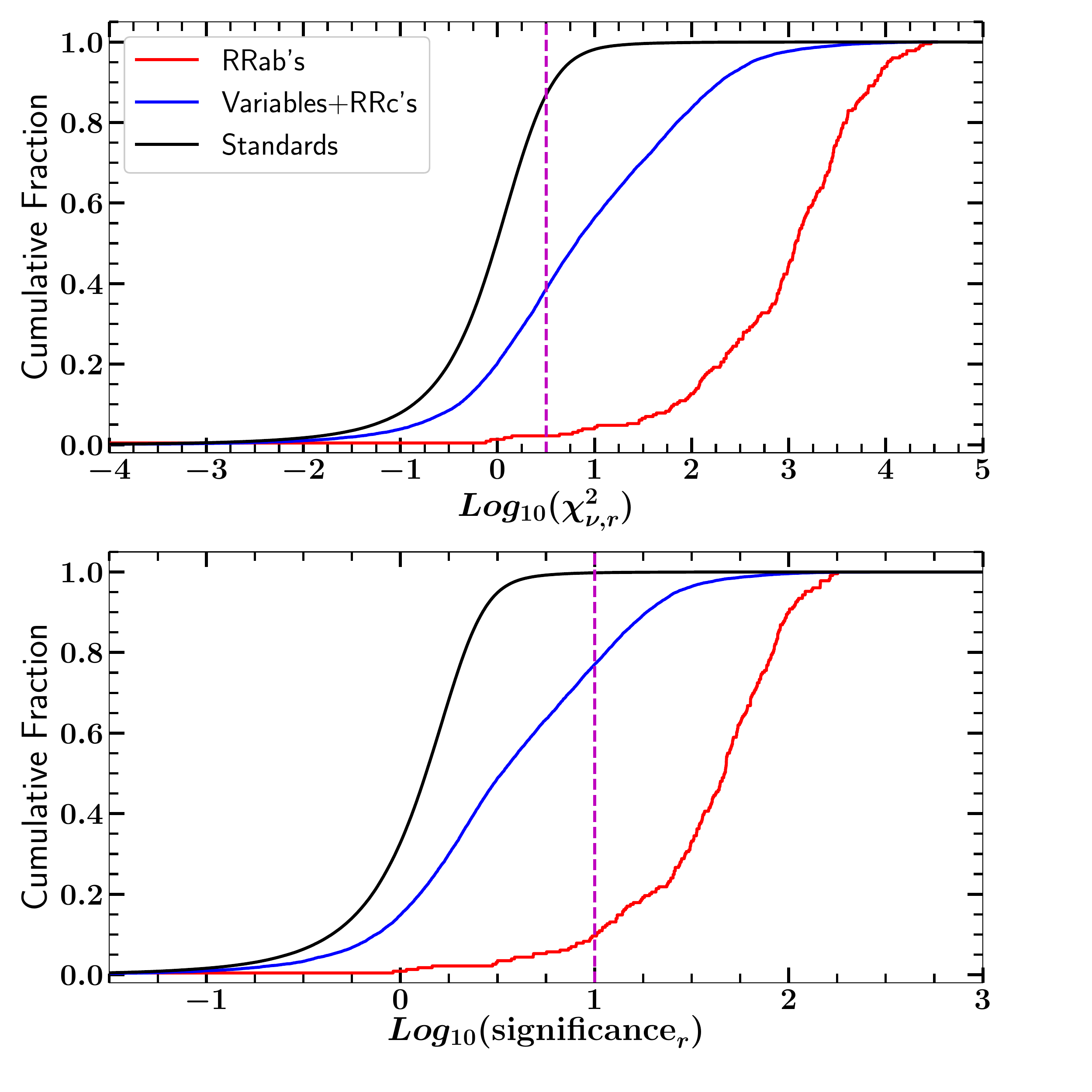}
  \caption{Initial variability metric values for previously identified objects in S82. \textit{Top:} Cumulative distribution of $\log_{10}(\chi_{\nu,r}^{2})$. The magenta dashed line denotes the chosen threshold value of 0.5; objects with larger values in any band are considered variable. \textit{Bottom:} Histogram of $\log_{10}($significance$_r)$. The vertical magenta dashed line denotes the chosen threshold value of $1.0$; objects with larger values in any band are considered variable. Note that although some real RRab are excluded by this cut, most of the non-RRab variables are also excluded.}
  \label{fig:init_cuts}
\end{figure}

As an example, we show the cumulative distributions of $\chi_{\nu,b}^{2}$ values and ``significance'' values for the cross-matched objects in the \textit{r} band in Figure \ref{fig:init_cuts}. For both metrics, the threshold values were chosen to minimize the number of non-variable stars that would be subject to subsequent analysis. Any objects that showed $\log_{10}(\chi_{\nu,b}^{2})\ge 0.5$ and significance $\ge1$ in any one of \textit{grizY}\footnote{Unlike the initial cuts we applied to the coadded catalog, we included the \textit{Y} band in these variability cuts because the \textit{Y} band values were weighted by the photometric uncertainties.} were kept for subsequent analysis. When these cuts were applied across all five filter bands, 234 $(\sim 98\%$) RRab, 57 ($\sim 98\%$) RRc, 5196 ($\sim 31\%$) variable, and 3004 ($\lesssim 0.05\%$) standard light curves from S82 met these criteria. These results for our training set are summarized in Table \ref{tab:training_set}. Over the entire survey region, approximately $\sim 7 \times 10^{5}$ light curves passed these variability cuts. We caution that passing this criterion is simply an initial cut and does not imply that these sources are truly variable.

\begin{deluxetable}{lrr}
	\tabletypesize{\footnotesize}
	\tablecaption{Training set of cross-matched S82 objects}
    \tablehead{
    \colhead{SDSS Label} & \colhead{Present in DES} & \colhead{Passed Cuts}}
	\startdata
    RRab & 238 & 234 \\
    RRc & 58 & 57 \\
    Variables & 16,752 & 5196 \\
	Standards & 641,710 & 3004 \\
	\enddata
    \tablecomments{Objects were originally identified in \citet{ivezic2007,sesar2010,suveges2012}.}
    \label{tab:training_set}
\end{deluxetable}

Despite these cuts, a small but non-negligible fraction of the objects identified as ``standard'' by \citet{ivezic2007} were still selected. It is possible that some of these objects are truly variable sources that did not display significant changes in the previous studies. Another possibility is erroneous photometry that, while rare, occurs sometimes in the Y3Q2 dataset due to incorrectly attributing observations from separate sources to one object or imperfect masking of observations obtained in very poor weather conditions. While these objects may have passed the initial variability cuts, their light curves fit the RRL template poorly, and most of them were rejected later in our analysis.
 
All of the selected light curves were corrected for extinction using reddening values from the maps of \citet{schlegel1998} multiplied by filter coefficients derived from the \citet{fitzpatrick1999dust} reddening law (for $R_{V}=3.1$) and the adjustments to the \citet{schlegel1998} map presented by \citet{schlafly2011} (see \S4.2 in \citet{abbott2018} for more details). We then used the extinction-corrected light curves as the input for our template fitting algorithm.

\section{Candidate Identification}
\label{sec:template}

\subsection{RR Lyrae Template}
\label{sec:template_description}

Our current work introduces a novel method of identifying RRab by fitting an empirically derived periodic model to the sparsely sampled multiband light curves. The model has the form:

\begin{equation}\label{eq:model}
m_{b}(t) = \mu + M_{b}(\omega) + a \gamma_{b}(\omega t + \phi) 
\end{equation}

\noindent where $m_{b}$ is the measured magnitude in a given band \textit{b} at a given time $t$, $\mu$ is the distance modulus, $M_{b}$ is the average absolute magnitude in that band, $\omega=1/P$ is the frequency of the variability (inverse of the period $P$), $a$ is the $g$-band amplitude (the amplitudes of the curves for the other bands are proportional to $a$), $\gamma_{b}$ is a periodic shape function, and $\phi$ is the phase. Only the four parameters $\mu,\ a,\ \omega,\ \phi$ are estimated during the fitting process while the forms of $\gamma_{b}$ and $M_{b}(\omega)$ are fixed. These were derived using well-sampled RRab light curves from S82 \citep{sesar2010} and shifted to adjust for slight differences between SDSS and DES filters. A more detailed description of the template construction is included in Appendix \ref{sec:appendix_model_assumptions}.

Using the same reasoning as \citet{sesar2017}, we chose to exclude RRc from our study because: a) their sinusoidal light curves are difficult to distinguish from light curves from other variable objects such as eclipsing binaries, b) their small amplitudes would make them difficult to identify in our sparse data, and c) searching over a larger period range to recover their short periods ($\sim$ 0.3d) would introduce additional common period aliases into our sample. While excluding RRc weakens our sample size for tracing substructure, it is not prohibitive since RRc are usually less numerous than RRab. Furthermore, this approach allowed us to use only one generalized RRab shape for our template instead of an ensemble of shapes as in \citet{sesar2017}. While they were able to recover a more diverse group of RRL by fitting multiple shapes, our approach makes our algorithm more computationally efficient.

\begin{figure}[!tb]
\includegraphics[width=0.5\textwidth]{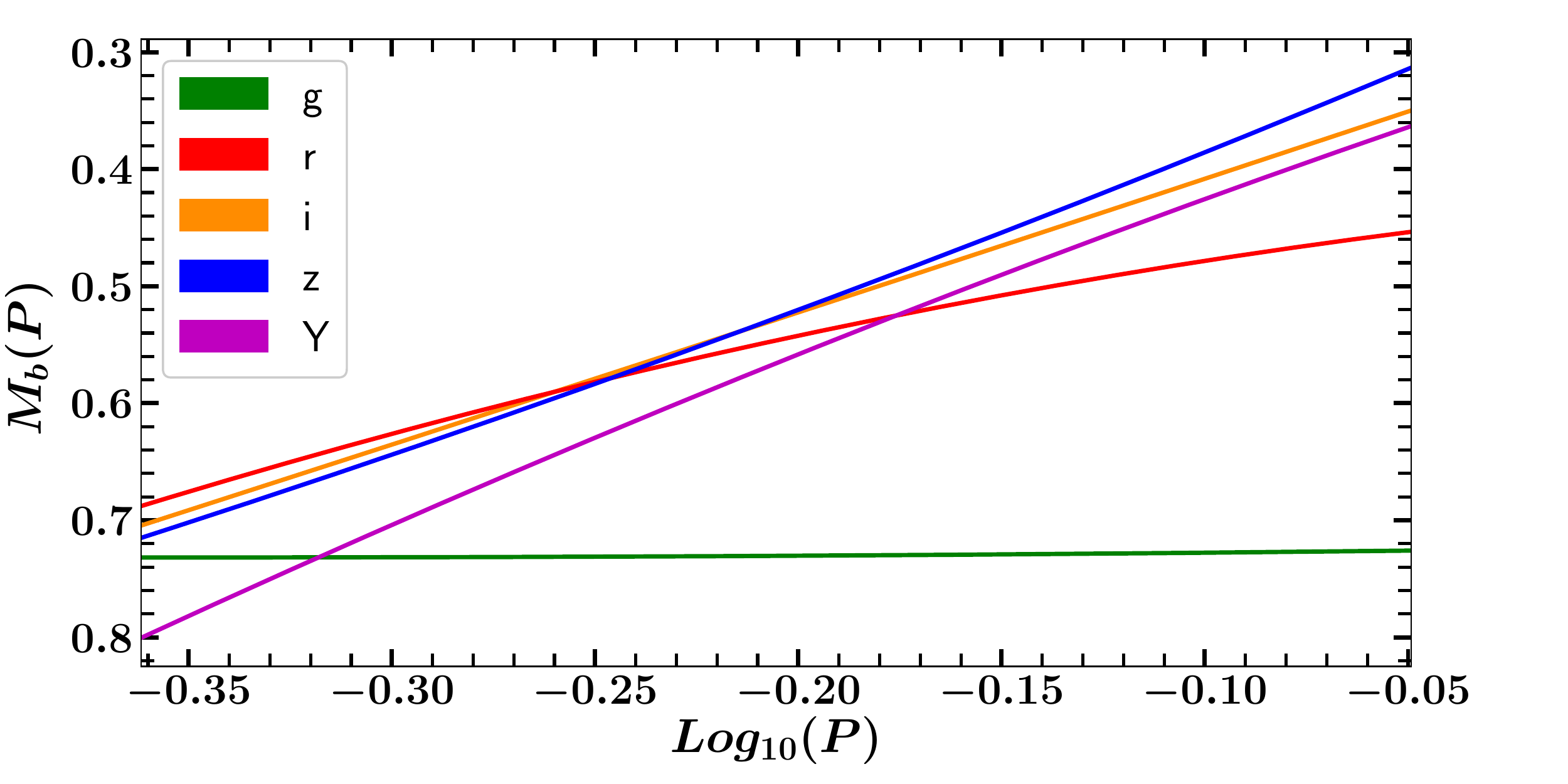}
\caption{$P-L$ relations used in our template fitting procedure. RRL have nearly constant absolute magnitudes in $g$ regardless of period. See \S\ref{sec:template_description} for more details.} 
\label{fig:absmag}
\end{figure}

The \textit{P-L-Z} relations were implemented in our template fitting procedure as \textit{P-L} relations evaluated at a starting metallicity of $[\mathrm{Fe/H}]=-$2.85. However, the values of these offsets between template filter curves were later adjusted to fit the S82 RRL light curves, so this metallicity should not be treated as the true value of the template. Hence, the metallicity value of our template is somewhat ambiguous. We discuss the effects of the metallicity further in \S\ref{sec:uncertainties}.  The final template \textit{P-L} relations are shown in Figure \ref{fig:absmag}. More details on how these parameters were derived are presented in Appendix \ref{sec:appendix_population_parameters}. 

\vfill
\subsection{Template Fitting}
\label{sec:template_fitting}

In this section, we include a high-level overview of the template fitting procedure. A more detailed description of this process can be found in Appendix \ref{sec:appendix_model_fitting}.

To fit the template to a light curve, the algorithm performs a grid search over a specified range of frequencies. To prevent misestimated and aliased periods outside of the true range of periods known for RRab, we restricted this range to values corresponding to periods of 0.44 days to 0.89 days following the period-amplitude relation shown in Figure 16 of \citet{sesar2010}. At each frequency gridpoint, the algorithm first calculates $M_{b}(\omega)$ (see Equation \ref{eq:model} and Figure \ref{fig:absmag}) and subtracts these values from the light curve magnitudes in the appropriate band. Then, with the frequency fixed, the model alternates between estimating the best-fitting $\phi$ using Newton's method, and ($a, \mu$) using weighted least squares. The ($\phi,a,\mu$) values which minimize the weighted Residual Sum of Squares (RSS) at each frequency gridpoint are chosen as the best-fitting parameters, and act as the starting point for the parameter search at the next gridpoint. The ($\omega,\phi,a,\mu$) at the global minimum RSS value over the entire frequency grid are chosen as the best-fitting parameters\footnote{In his study of Cepheid variables, \citet{stetson1996} also developed a template fitting method based on least squares. However, instead of using a string-length minimization technique in a single band, we use all the bands simultaneously and used a fixed shape instead of a family of derived template curves calculated for each trial period.}.

\begin{figure}[!t]
\includegraphics[width=0.5\textwidth]{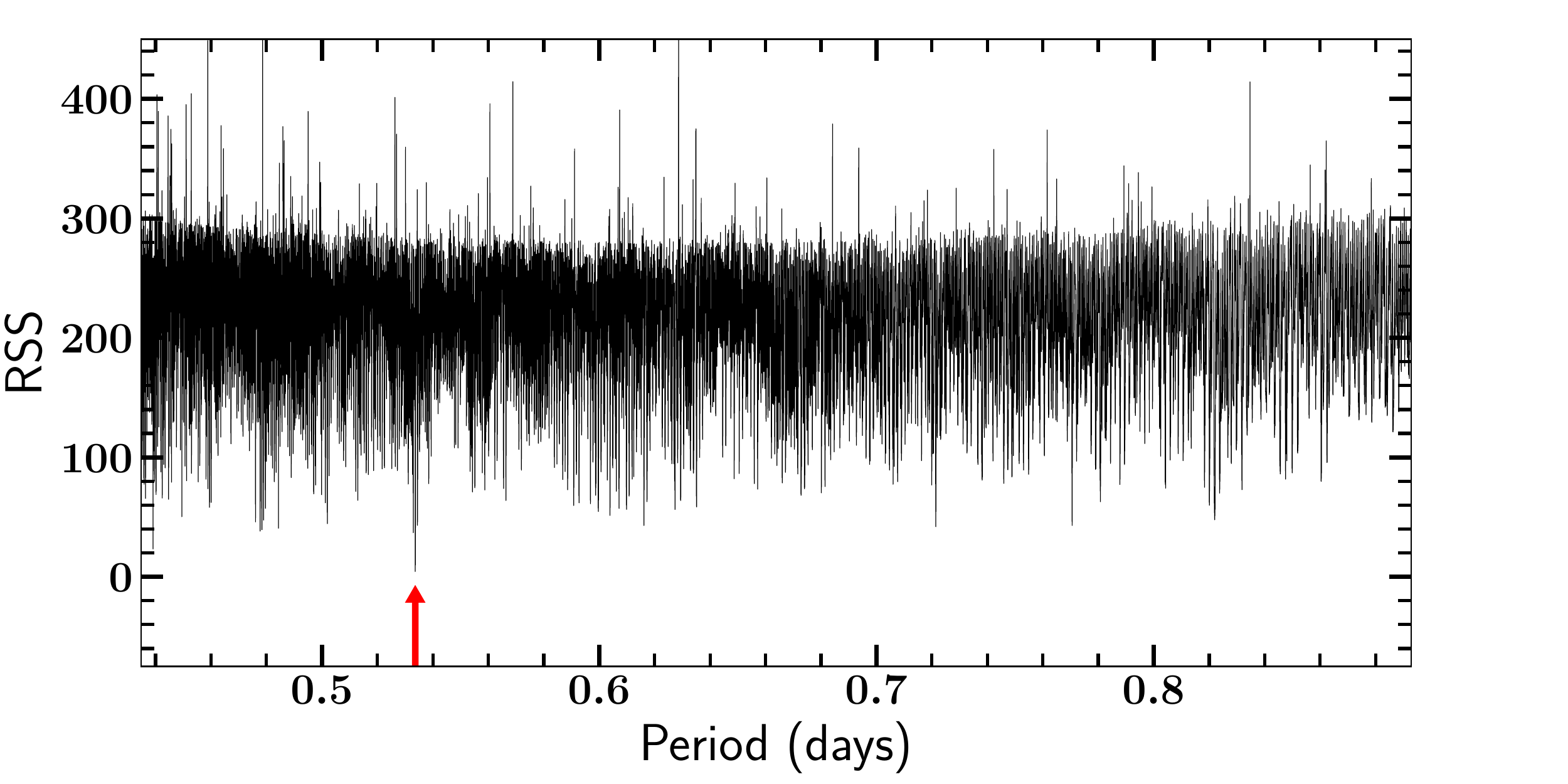}
\caption{Residual Sum of Squares (RSS) curve for an RRab originally discovered by \citet{sesar2010}. The red arrow denotes the global minimum of the RSS which corresponds to the true period of 0.5336 days.}
\label{fig:rss}
\end{figure}

A major strength of this algorithm is that the template shape is fit simultaneously to the light curve data in all five bands combined. Since there are only four free parameters that must be fit for the entire light curve, unique solutions can be found for sparse light curves with very few measurements in any single band. An example RSS curve for an RRab from the labeled training set is shown in Figure \ref{fig:rss}. Because the local minima in the RSS curve are sometimes very close in value, we include the best-fitting values for the top three minima of RSS in our data products for completeness, but do not discuss the results of the 2nd and 3rd minima further in this work. 

\begin{figure}[!tbh]
\includegraphics[width=0.5\textwidth]{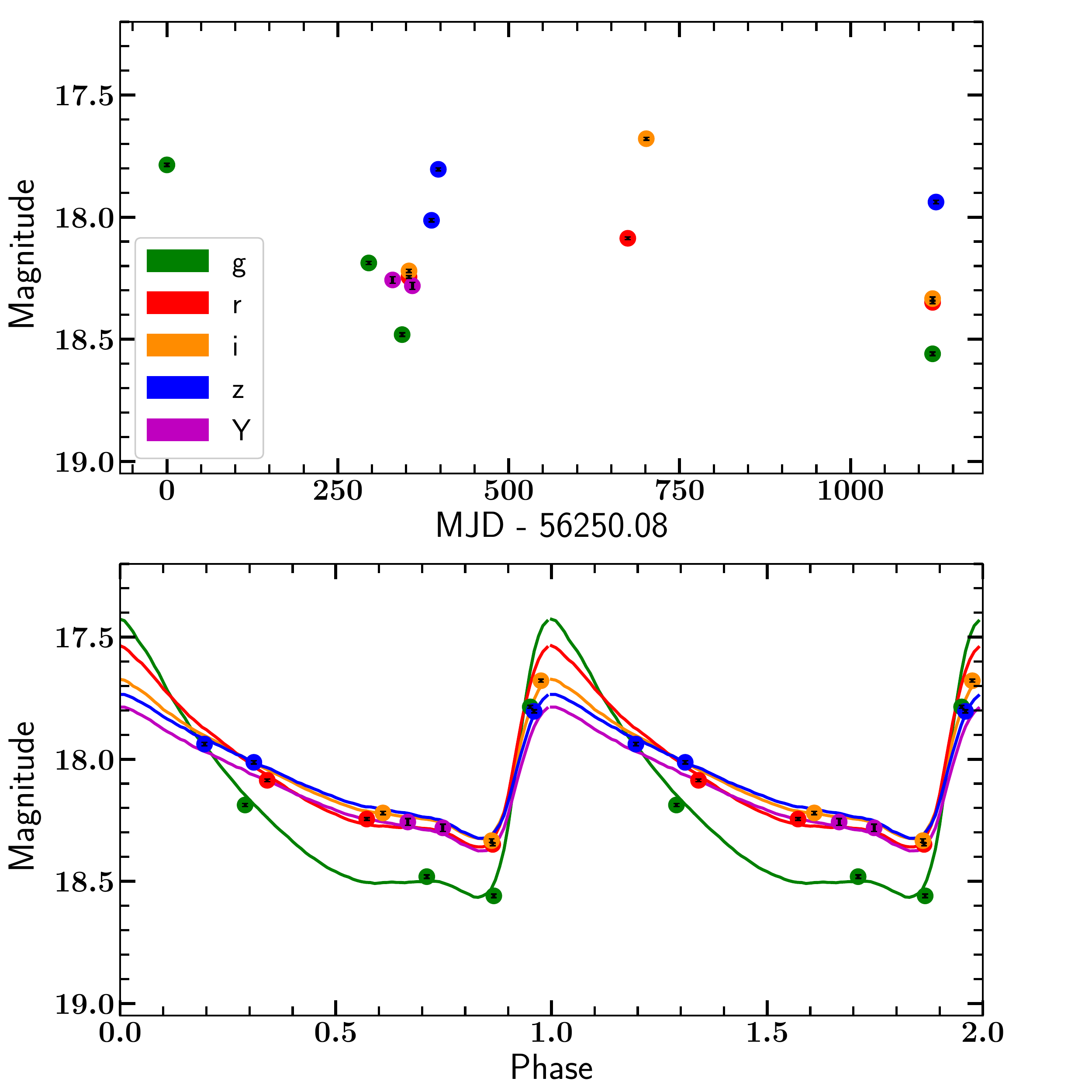}
\caption{\textit{Top}: Poorly sampled DES light curve of an RRab originally discovered by \citet{sesar2010} (same star as Figure \ref{fig:rss}). 
\textit{Bottom}: Phased light curve of the same source with the period correctly estimated by our algorithm.
\textit{Note:} Photometric uncertainties are smaller than the plotting symbols.} 
\label{fig:template}
\end{figure}

Our algorithm is also effective at estimating distances. At the best fit parameter estimates, the template fitting algorithm correctly estimated $\sim$81\% of the S82 RRL distances to within 3\% of the values obtained by \citet{sesar2010} and \citet{suveges2012} (if available) with an overall standard deviation of 2.8\% (see \S\ref{sec:uncertainties} for an extended discussion of the uncertainties in distance modulus). The accuracy of the template estimates of both the period and the distance for the training set of RRab light curves is summarized in Table \ref{tab:accuracy}).

\begin{deluxetable}{cccc}
	\tabletypesize{\footnotesize}
	\tablecaption{Period and distance estimation accuracy} 
    \tablehead{
      \multicolumn{1}{c}{Parameter} &
      \multicolumn{2}{c}{\% of RRab within} & 
      \multicolumn{1}{c}{$\sigma_{\mathrm{parameter}}$}\\ 
      &
      \multicolumn{1}{c}{\ \ \ \ \ \ 1\%} & 
      \multicolumn{1}{c}{\ \ \ \ \ \ 3\% \ \ \ \ \ \ } & 
    }
	\startdata
    $\Delta P / P_{\mathrm{prev}}$ & 88.89 & 89.74 & 6.81\% \\
	$\Delta D / D_{\mathrm{prev}}$ & 44.64 & 81.11 & 2.83\% \\
	\enddata
    \tablecomments{$P_{\mathrm{prev}}$ and $D_{\mathrm{prev}}$ are the values reported in \citet{sesar2010} and \citet{suveges2012}.}
    \label{tab:accuracy}
\end{deluxetable}

Our algorithm is computationally efficient and only takes $\sim 3-5$ minutes per light curve on an Intel Xeon E5420 processor. The template fitting code returned the estimated parameters of the top three best-fitting templates as well as the features used in the random forest classification detailed in \S\ref{sec:rf} and Table \ref{tab:features}. The computation time for fitting the template and calculating features for $\sim7\times 10^5$ light curves was $\sim44$K CPU hours. For comparison to a similar analysis, our algorithm is $>9\times$ faster than the template fitting methods used by \citet{sesar2017}, which required $\sim 30$ minutes per star. 

There are several other well-documented methods available in the literature for identifying RRL in multiband data (e.g. \citealt{vanderplas2015}, \citealt{hernitschek2016} and \citealt{sesar2017}), which yield excellent results for data sets with an average of 35 or more observations per light curve. We present this algorithm as an alternative to these other methods for especially sparse multiband data sets (see \S\ref{sec:simresults} for a discussion of the observational limitations.) The template and the associated fitting code are available at \url{https://github.com/longjp/rr-templates} and further described in Appendix \ref{sec:appendix_template_code}. 

\subsection{Feature Selection}
\label{sec:features}

While it is possible to identify RRL by visually inspecting their phase-folded light curves, the sheer volume of light curves in our data set makes this classification method unfeasible. Instead, we computed numerical features to describe the behavior of the light curves. To assess the specific parameter space occupied by RRab, we compiled a training set consisting of all the cross-matched labeled objects from S82 which passed the initial variability cuts (discussed in \S\ref{sec:varcuts}). This left us with a training set of 234 RRab, 57 RRc, 5196 other variable objects, and 3004 ``standard'' sources.  Since we only aimed to identify RRab, we chose a simple identification scheme with two classes: RRab and non-RRab. This resulted in an RRab class with 234 objects and a non-RRab class with 8257 objects.

\begin{deluxetable*}{llr}
	\tabletypesize{\footnotesize}
	\tablecaption{Random Forest Features}
	\tablehead{\colhead{Feature Name} & \colhead{Description} & \colhead{Importance}}
	\decimals
	\startdata
    lchi\_med & median $\mathrm{log_{10}}(\chi_{\nu,b}^{2})$ value across all bands & 0.2232\\
    amp\_rss\_0 & Best-fitting Amplitude divided by the best-fitting RSS/dof & 0.1942\\
    f\_dist1\_0 & Closest distance of best-fitting period/amplitude to Oosterhoff I relation from \citet{sesar2010} &  0.1591\\
	rss\_dof\_0 & RSS of the best-fitting template per degree of freedom & 0.1571 \\
	f\_dist2\_0 & Closest distance of best-fit period/amplitude to \citet{sesar2010} curve dividing the Oosterhoff I and II groups & 0.1025\\
	amp\_0 & Best-fitting amplitude & 0.0792\\
    rss\_lchi\_med & RSS of the best-fitting template divided by the median $\mathrm{log_{10}}(\chi_{\nu,b}^{2})$ value across all bands & 0.0502 \\
	$\kappa$\_0 & Best-fitting von Mises-Fisher concentration parameter of phases in the folded light curve & 0.0345\\
	\enddata
	\tablecomments{All of these feature values for each candidate are included in the electronic version of Table \ref{tab:candidates}.}
    \label{tab:features}
\end{deluxetable*}

With the goal of identifying RRab, we chose features which were motivated by how well the light curves fit the RRab template and other observed properties of RRab. As demonstrated in Figure \ref{fig:init_cuts}, RRab have relatively large $\mathrm{log_{10}}(\chi_{\nu,b}^{2})$ compared to most of the other objects in our sample. So, to quantify the base variability of the light curve while ignoring spurious signals or missing observations in any particular band, we included the median of this value calculated across all five filters as a feature. Additionally, most true RRab should fit the template with small residuals, so we quantified the quality of the best template fits using the RSS per degree of freedom, RSS$_{\mathrm{dof}} = \mathrm{RSS}/(N_{obs} - 4)$. Then, to amplify the separation provided by these two characteristics, we divided the RSS$_{\mathrm{dof}}$ by the median $\mathrm{log_{10}}(\chi_{\nu,b}^{2})$ to form another feature. 

To take advantage of the distinctive amplitude ranges of RRab, we also selected the amplitude of the best-fitting template as a feature. We then created a new feature by dividing this amplitude by the RSS$_{\mathrm{dof}}$, expecting that the large amplitudes and excellent template fits of RRab would clearly distinguish them from other objects. We can take advantage of these amplitudes again by evaluating how closely each object matches the observational trends shown by RRab in the first two Oosterhoff groups (see the introduction for a brief description). To measure how closely the objects' estimated template parameters matched these trends for RRab, we calculated the distance of the object in period-amplitude space from the Oosterhoff I relation measured in \citet{sesar2010} and their shifted curve which separates the Oosterhoff I and II populations (see their Figure 16 and our Figure \ref{fig:bailey}.) 

Our last feature attempted to quantify the phase distribution of the observations in each light curve. Period-finding algorithms often recover periods at common aliases, sometimes resulting in light curves with many of their observations clustered near a particular phase. Because light curve phases are periodic, the two-dimensional case of the von Mises-Fisher distribution \citep{fisher1953,jupp1989} is a good approximation. This distribution can be written as: 

\begin{equation}
f(x) = \frac{e^{\kappa cos(x-\mu)}}{2\pi I_{0}(\kappa)}
\label{eq:vonmises}
\end{equation}

\pagebreak

\noindent where $\kappa$ is the concentration parameter, $\mu$ is the mean, and $I_{0}(\kappa)$ is the modified Bessel function of the first kind at order 0. The von-Mises Fisher distribution is akin to a Gaussian distribution wrapped around a circle, where the $\kappa$ parameter is analogous to the inverse of the variance (see \S2.2 in \citealt{sra2016} for a more detailed description). We calculated this concentration parameter $\kappa$ for each light curve folded over the best-fitting template period. Light curves with observations highly clustered in phase will have very large values of $\kappa$, aiding in the rejection of objects with aliased periods.

\begin{figure}[]
\includegraphics[width=0.5\textwidth]{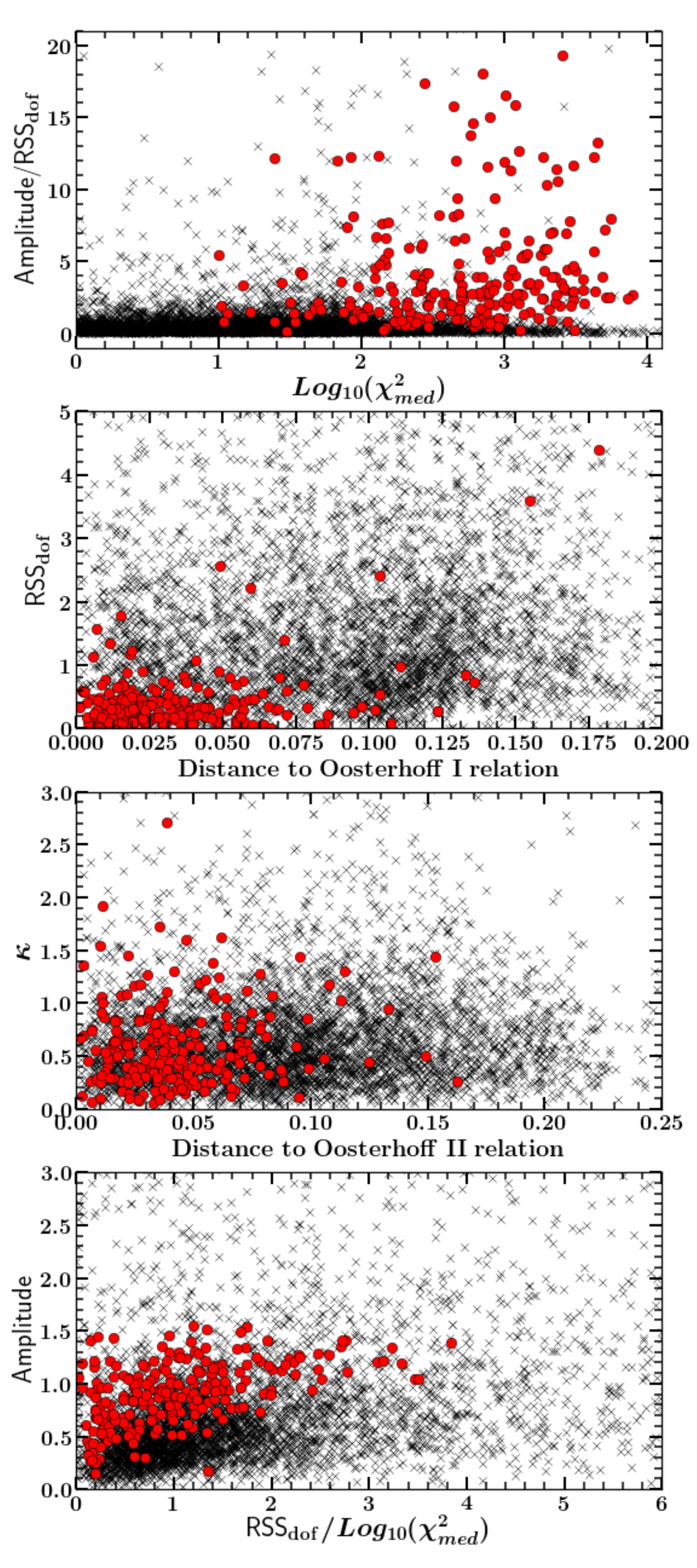}
\caption{Features used to identify RRab plotted for the training set. Red points denote previously identified RRab while black X's are non-RRab. While the RRab clearly occupy a specific region in this feature space, they are not linearly separable.}
\label{fig:top_features}
\end{figure}

Although our choice of classifier is generally robust against non-informative features, we limited our features to these to make the classifier results easier to interpret. The features are summarized and ranked by their importance, or how much they contributed to splitting the data across all the decision trees in our classifier\footnote{See \S 1.11.2.5 in the {\tt scikit-learn} documentation for more details.}, in Table \ref{tab:features}. These features are shown in Figure~\ref{fig:top_features}. The development of additional features to further separate the classes will be explored in future work.

\subsection{Random Forest Classifier}
\label{sec:rf}

To identify likely RRab automatedly and consistently, we trained a random forest classifier \citep{amit1997,breiman2001} using these features. The random forest is a machine learning algorithm that predicts classes for data by combining results from a ``forest" of decision trees. Each decision tree consists of a series of nodes where the data is split into subgroups based on the values of a random subset of their features, or characteristics. Before the random forest can make accurate predictions, it must be trained to recognize the trends in features that correspond to different classes. Thus, one needs a labeled training set to build the random forest. Each decision tree uses the labels to build a series of boundaries in feature space that divide the data into their correct classes. Once trained, the random forest algorithm assigns a score to unlabeled data based on the proportion of trees that identify them as a particular class. Random forest classifiers have been extremely successful in variable star classification (see \citealt{richards2011} for a comparison with other machine learning techniques), even in the case of sparsely sampled Pan-STARRS PS1 light curves \citep{hernitschek2016}. Thus, the random forest was a natural choice of classifier for this study.

\begin{deluxetable}{llr}
	\tabletypesize{\footnotesize}
	\tablecaption{Estimated RRab selection \\ \qquad \qquad purity \& completeness} 
    \tablehead{
    \colhead{Score Threshold} & \colhead{Purity} & \colhead{Completeness}}
	\startdata
    0.00 & 0.043 & 0.983 \\
	0.05 & 0.402 & 0.928 \\
	0.10 & 0.567 & 0.886 \\
	0.15 & 0.661 & 0.852 \\
	0.20 & 0.727 & 0.840 \\
	0.25 & 0.773 & 0.819 \\
	0.30 & 0.815 & 0.798 \\
	0.35 & 0.845 & 0.756 \\
	0.40 & 0.877 & 0.722 \\
	0.45 & 0.896 & 0.693 \\
	0.50 & 0.922 & 0.651 \\
	0.55 & 0.955 & 0.630 \\
	0.60 & 0.953 & 0.596 \\
	0.65 & 0.956 & 0.554 \\
	0.70 & 0.968 & 0.525 \\
	0.75 & 0.973 & 0.470 \\
	0.80 & 0.971 & 0.432 \\
	0.85 & 0.988 & 0.348 \\
	0.90 & 0.982 & 0.231 \\
	0.95 & 1.000 & 0.071 \\
	\enddata
    \tablecomments{The full version of this table is available in the online data products.}
    \label{tab:puritycompleteness}
\end{deluxetable}

We created the classifier using the {\tt RandomForest} package available in {\tt scikit-learn} \citep{scikit-learn}. To prevent overfitting to our small training set and ensure repeatability, the classifier contained 500 trees with a maximum depth of 5, and used a random seed of 10.

\begin{figure}
\includegraphics[width=0.5\textwidth]{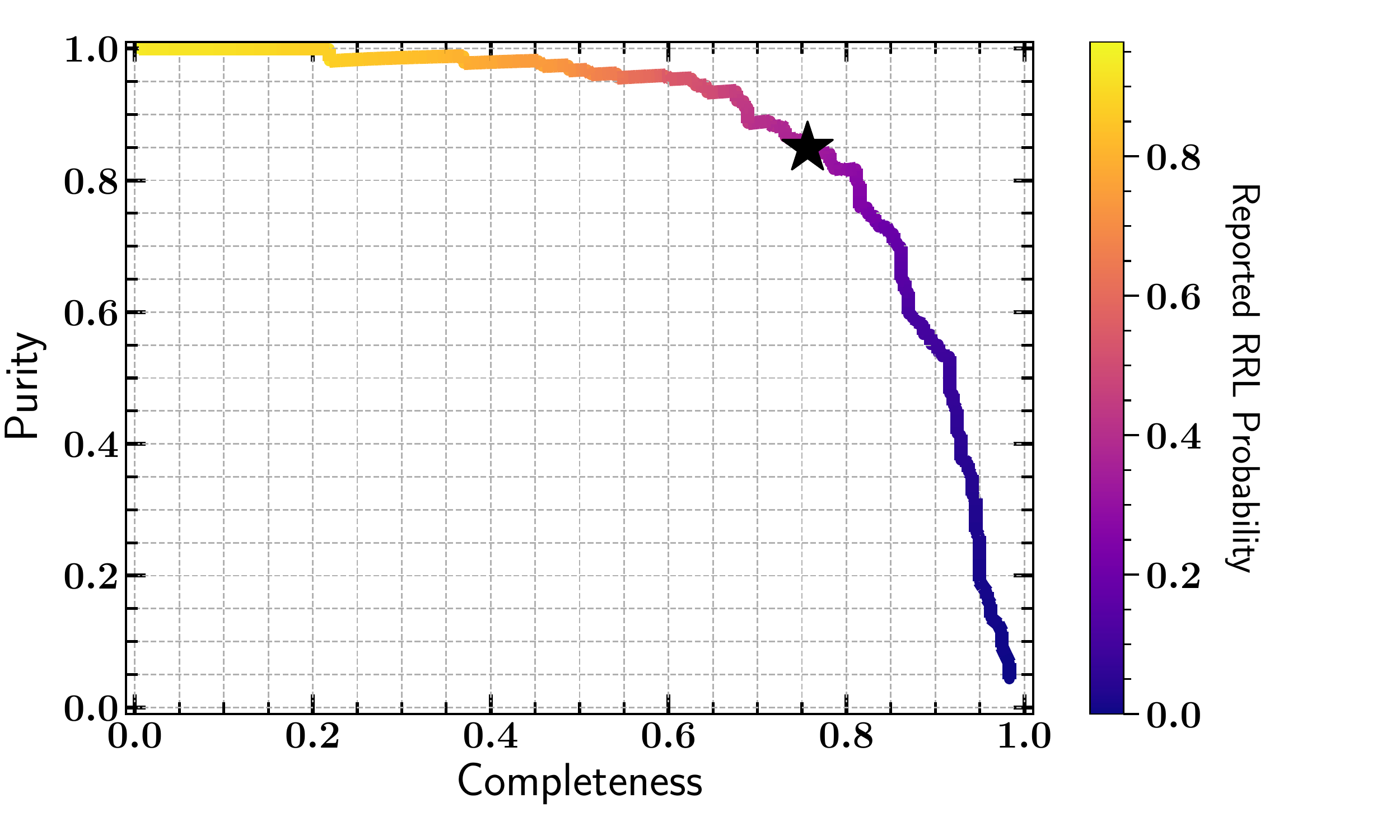}
\caption{Purity/Completeness curve for the random forest classifier trained on cross-matched objects in S82. The black star denotes a classifier-reported score of 0.35, where the purity is $\sim$ 85\% and the completeness is $\sim$ 76\%. The area under the curve is 0.864.}
\label{fig:roc}
\end{figure}

We assessed the performance of our classifier by estimating the purity (the percentage of objects classified as RRab that were truly so) and the completeness (the percentage of real RRab that were identified as such) as a function of the class score reported by the random forest. The purity and completeness were estimated using a five-fold cross validation technique, where the data were divided into five test groups and classified based on a classifier trained on the other four groups. The classifier correctly identified 190 of the 234 RRab used to train the random forest as such with a score threshold of $\ge 35\%$. As shown in Figure \ref{fig:roc}, defining RRab as objects with a score $\ge 35\%$ yields a purity of 85\% and a completeness of 76\%. A common metric used to assess the performance of a classifier is the area under the curve (AUC) shown in Figure \ref{fig:roc}, which we find to be 0.864. The purity and completeness calculated at other score thresholds are listed in Table \ref{tab:puritycompleteness}. We include all other objects with lower scores in our catalog so that other score thresholds can be specified by interested readers. Although an incorrect period estimate led to a worse RSS value for the fit, some RRab with incorrect period estimates were still correctly identified as such by the classifier. 

Since our training set was mostly composed of nearby RRab confined to a small region in the survey footprint, it is imperative to assess our classifier performance using other samples of RRL. To this end, we cross-matched our sample with external surveys in \S\ref{sec:externalcats} and applied our method to simulated RRab light curves at fainter magnitudes in \S\ref{sec:simresults}. 

\section{Catalog Description}
\label{sec:catalog}
\subsection{Visual Validation}
\label{sec:visual_validation}

\begin{figure*}
	\includegraphics[width=1\textwidth]{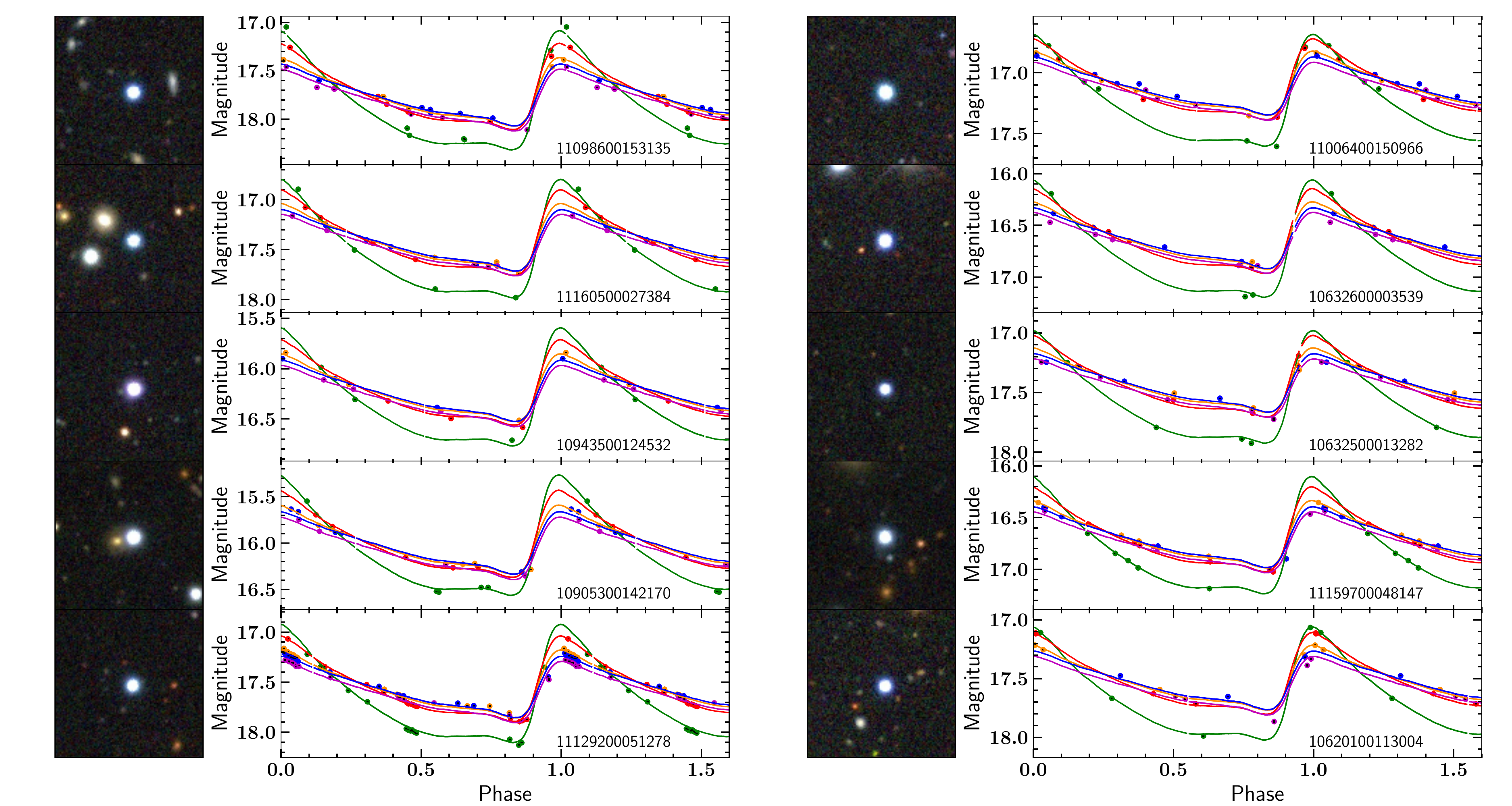}
	\caption{Sample of DES coadded images and representative light curves of visually accepted RRL candidates with classifier scores exceeding 0.94, labeled with their Y3Q2 ID number. The observations and templates are colored by filter using the same convention as Figure \ref{fig:template}.}
    \label{fig:real_highprob}
\end{figure*}

\begin{figure*}
	\includegraphics[width=1\textwidth]{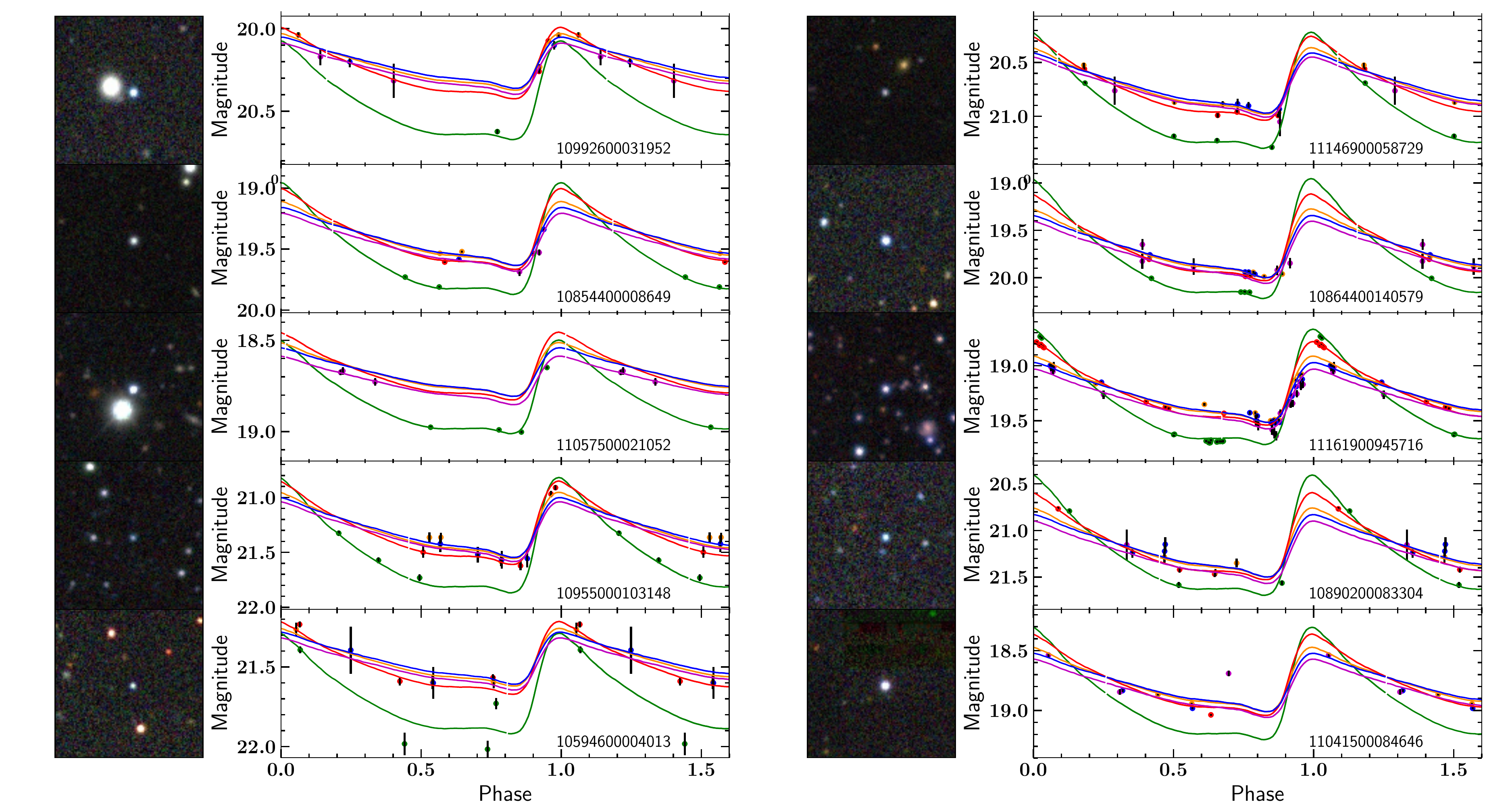}
	\caption{Sample of DES coadded images and representative light curves of visually accepted RRL candidates with classifier scores below 0.36, labeled with their Y3Q2 ID number. The observations and templates are colored by filter using the same convention as Figure \ref{fig:template}.}
    \label{fig:real_lowprob}
\end{figure*}

\begin{figure*}[]
	\includegraphics[width=1\textwidth]{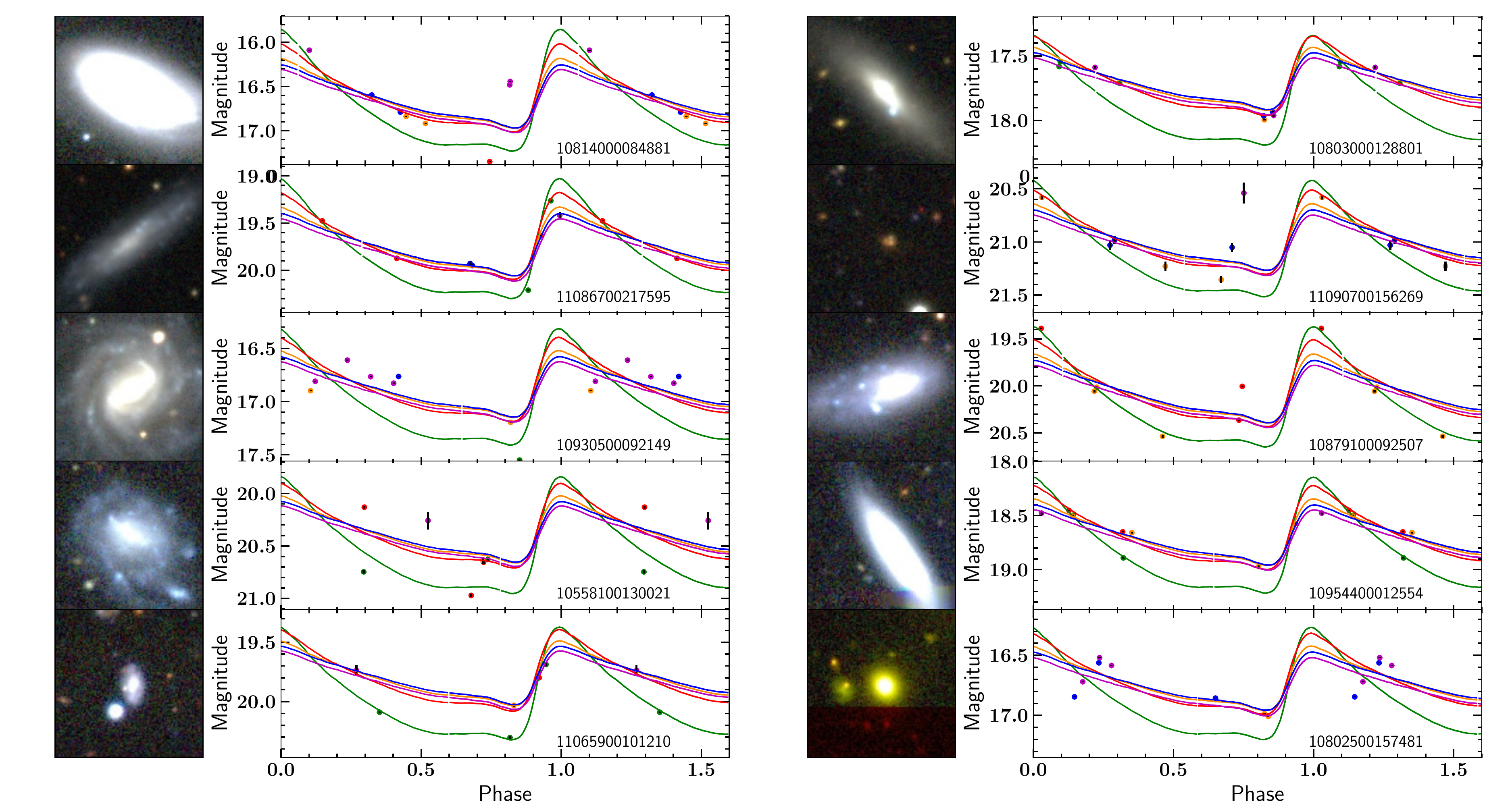}
	\caption{Sample of DES coadded images and representative light curves of visually-rejected candidates (extended sources or possible supernova) that passed the classifier score threshold, labeled with their Y3Q2 ID number. The observations and templates are colored by filter using the same convention as Figure \ref{fig:template}.}
    \label{fig:fake_galaxy}
\end{figure*}

\begin{deluxetable*}{lrrrrrrrrrrr}
	\tabletypesize{\footnotesize}
	\tablecaption{DES RRab Candidates}
	\tablehead{\colhead{DES Y3Q2 ID} & \colhead{$\alpha$} & \colhead{$\delta$} & \colhead{$\langle g \rangle$} & \colhead{$\langle r \rangle$} & \colhead{$\langle i \rangle$} & \colhead{$\langle z \rangle$} & \colhead{$\langle Y \rangle$} & \colhead{$P$} & \colhead{$A_{g}$} & \colhead{$\mu$} & \colhead{$p$}\\
    \colhead{} & \multicolumn{2}{c}{[deg, J2000]} & \multicolumn{5}{c}{[mag]} & \colhead{[d]} & \multicolumn{2}{c}{[mag]}}
    \decimals
    \startdata
    11136400113264 & 0.000107 & -59.559187 & 17.657 & 17.500 & 17.547 & 17.535 & 17.588 & 0.6415 & 0.551 & 16.90 & 0.436 \\
    10646400013129 & 0.013042 & -2.430057 & 17.234 & 17.052 & 17.099 & 17.103 & 17.252 & 0.4938 & 1.151 & 16.66 & 0.733 \\
    11004800140792 & 0.131498 & -41.482218 & 19.229 & 19.096 & 19.125 & 19.195 & 19.145 & 0.5893 & 1.137 & 18.69 & 0.568 \\
    11108800089990 & 0.134204 & -54.295118 & 15.754 & 15.625 & 15.557 & 15.554 & 15.571 & 0.6698 & 0.469 & 14.97 & 0.659 \\
    10595200009863 & 0.283437 & 1.178535 & 17.986 & 17.922 & 17.908 & 17.868 & 17.869 & 0.5480 & 1.033 & 16.96 & 0.886 \\
    \enddata
	\tablecomments{DES Y3Q2 ID: DES Y3Q2 {\sc quick\_object\_id} number. $\alpha$: Right Ascension. $\delta$: Declination. $\langle grizY \rangle$: Mean extinction-corrected magnitude. $P$: Best-fit period. $A_{g}$: Best-fit amplitude in DES $g$. $\mu$: Best-fit distance modulus. $p$: RRab score assigned by the classifier. The full version of this catalog, including feature values and cross matching information, is available in the online data products at \url{https://des.ncsa.illinois.edu/releases/other/y3-rrl}.}
    \label{tab:candidates}
\end{deluxetable*}

We applied the classifier to the $\sim 7\times 10^{5}$ objects with template fits and found 8026 RRL candidates with a score $\ge 0.35$. Although most of our candidates were indeed RRL found in other surveys, there were still some non-stellar interlopers in the sample due to the lenient initial cuts on the shape of the  photometric point-spread function (\S\ref{sec:selection}). Thus, we visually inspected all RRab candidate light curves and their DR1 coadded images. After visually validating the candidates and removing any objects with non-RRL classifications in the Simbad database \citep{simbad2000}, 1786 objects were discarded and 5783 RRL candidates remained in the catalog, with the rest too ambiguous to confirm. A sample of visually verified candidates with high ($p > 0.94$) and low ($p < 0.36$) classifier scores are shown in Figures \ref{fig:real_highprob} and \ref{fig:real_lowprob}, respectively. The most typical contaminants in the sample were extended sources. Given our lenient selection criteria described in \S\ref{sec:selection}, it is not surprising that some of these objects made it into our final sample. Examples of candidates that were rejected by visual inspection as being extended sources are shown in Figure \ref{fig:fake_galaxy}. The full catalog of candidates, their best-fit parameters, and their features are available at \url{https://des.ncsa.illinois.edu/releases/other/y3-rrl}. A sample view of this catalog is shown in Table \ref{tab:candidates}. Appendix \ref{sec:appendix_data_products} contains a detailed description of these data products.

Although the light curves have been visually inspected, further photometric observations of some extremely poorly sampled candidate light curves would be useful to confirm their classification. One may wish to remove these more uncertain candidates from their analysis by only considering objects with a larger minimum number of observations. Some of these candidates have gaps in observations near their maximum and minimum brightness, providing poor constraints on their estimated amplitudes and mean magnitudes. Therefore, we assigned a flag to each object based on how its phase-folded light curve is sampled. An object with fewer than two observations near its minimum brightness ($0.55\leq \mathrm{phase} <0.87$, which we chose to encompass both the near-constant portion of the light curve where other authors chose their minimum phase \citep[e.g.,][]{vivas2017} and the 10\% quantile of template magnitudes) will receive a ``flag\_minmax" value of 1, while an object with $<2$ observations near its maximum brightness ($0.96\leq \mathrm{phase} <1$ or $0\leq \mathrm{phase} <0.05$ corresponding to the 90\% quantile in template magnitudes) receives a ``flag\_minmax" value of 2. Objects missing observations near both of these receive a flag value of 3.

\begin{figure}[!bth]
\includegraphics[width=0.5\textwidth]{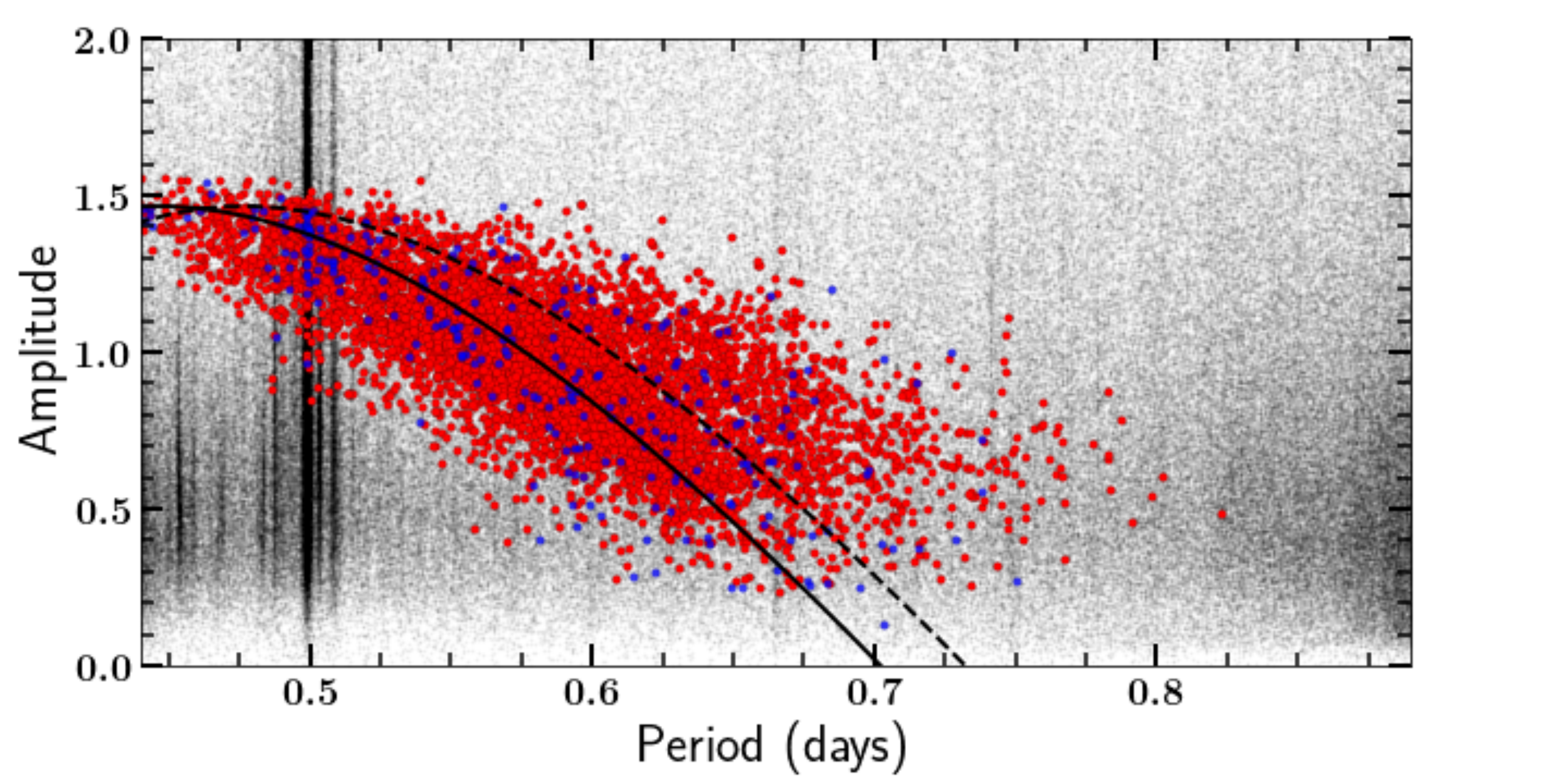}
	\caption{Bailey diagram of template-estimated amplitudes and periods for objects that passed the initial variability cuts in black and visually accepted RRab identified by our classifier in red. Ambiguous candidates that could not be visually accepted are plotted in blue. We overplot the Oosterhoff I relation and the curve dividing the Oosterhoff I and II populations from \citet{sesar2010} in black solid and dashed lines, respectively. The abundance of objects with periods of $P=0.5$ day denotes a common alias of the 1 day rotation period of the Earth.}
    \label{fig:bailey}
\end{figure}

Figure \ref{fig:bailey} shows a Bailey (period-amplitude) diagram of the candidates rejected by the classifier or our visual checks plotted in black, visually accepted candidates shown in red, and ambiguous candidates in blue. We plot the Oosterhoff I \citep{oosterhoff1939} relation and the curve dividing groups I and II from \citep{sesar2010} which we used in the calculation of our features in solid and dashed black lines. Many of these ambiguous candidates are likely RRab, but cannot be classified as such with high confidence in this work. Due to the sparse nature of our observations, we cannot detect amplitude modulations such as those arising from the Bla{\v z}ko effect \citep{blazhko1907}, although we did recover five out of twelve Bla{\v z}ko RRL previously identified in the Catalina Surveys \citep{drake2014,drake2017}.

\subsection{Comparison with Overlapping Catalogs}\label{sec:externalcats}

\begin{figure*}
\includegraphics[width=1\textwidth]{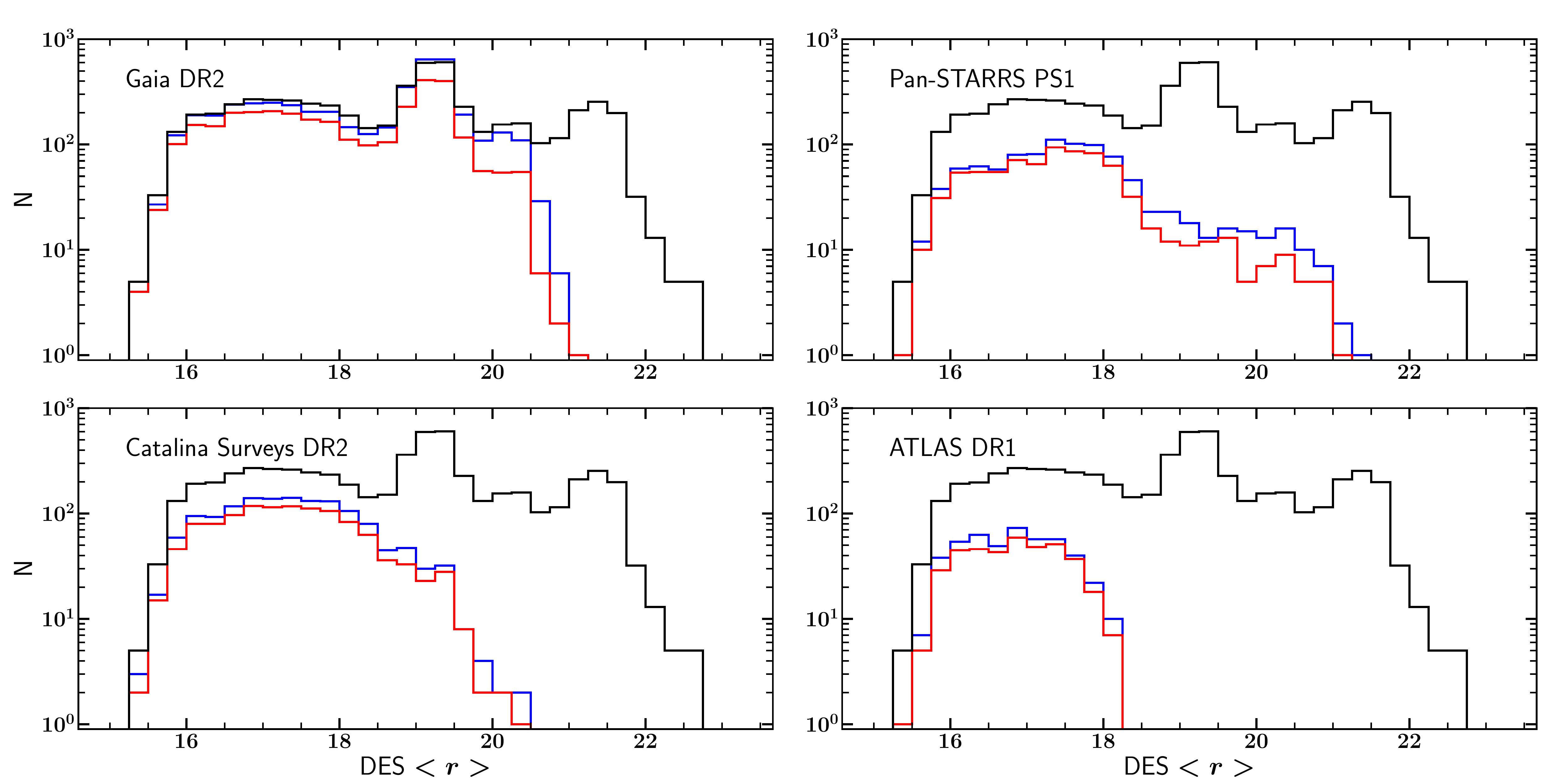}
\caption{Histograms of magnitudes of RRab stars from external catalogs cross-matched with our DES initial stellar catalog, as a function of the extinction-corrected weighted average coadded DES $r$ magnitude. \textit{Top left: Gaia} DR2. \textit{Top right}: Pan-STARRS PS1. \textit{Bottom left}: Catalina Surveys DR2. \textit{Bottom right:} ATLAS. Blue curves show the RRab from each catalog that were present in our sample before applying any cuts, while red curves show those that were identified as RRab in our analysis. Black curves show the distribution of DES RRab candidates and are the same in all panels. The overdensities at $r\approx 18.8$ and $r\approx 21.2$ correspond to the LMC outskirts and the Fornax dSph. Our catalog is deeper than the others by $\sim 1, 1, 2$ and $4.5$~mag, respectively.}
\label{fig:externalcats}
\end{figure*}

The DES footprint has significant overlap with other surveys, such as {\it Gaia} \citep{gaia2016}, Pan-STARRS \citep{chambers2016panstarrs}, the Catalina Surveys \citep[][]{drake2009}, and the Asteroid Terrestrial-impact Last Alert System \citep[ATLAS:][]{tonry2018}. We used our cross-matches with these external RRab catalogs to independently assess the performance of our algorithm at the different magnitude ranges probed by these surveys. We used the \texttt{SkyCoord} package in {\tt astropy} \citep{astropy2018} to select matches within $1\arcsec$ of DES objects while removing duplicates. Details for each individual survey are presented in the following paragraphs and summarized in Table \ref{tab:external}, while Figure~\ref{fig:externalcats} shows the respective overlaps with DES.

\begin{deluxetable*}{lrrrrrrrr}
	\tabletypesize{\footnotesize}
	\tablecaption{Description of Selected External RRL Catalogs and their Overlap with DES}
	\tablehead{\multicolumn{1}{c}{\multirow{2}{*}{Survey}} & \multicolumn{1}{c}{Area} & \multicolumn{1}{c}{\multirow{2}{*}{Filters}} & \multicolumn{1}{c}{\multirow{2}{*}{Depth}} & \multicolumn{1}{c}{Observational} & \multicolumn{1}{c}{\multirow{2}{*}{$N_{\rm obs}$}} & \multicolumn{3}{c}{RRab}\\
    \multicolumn{1}{c}{} & \multicolumn{1}{c}{[sq deg]} & \multicolumn{1}{c}{} & \multicolumn{1}{c}{} & \multicolumn{1}{c}{cadence} & \multicolumn{1}{c}{} & \multicolumn{1}{c}{total} & \multicolumn{1}{c}{in DES} & \multicolumn{1}{c}{\% found}} 
	\startdata
    SDSS stripe 82 & $\sim$300 & $ugriz$ & $g$$\le$21 & most observed every 2 days & 70-90 & 447 & 238 & 75 \\
    Gaia DR2 & all sky & $G_{BP},G,G_{RP}$ & $G$$\sim$21 & uneven, follows Gaia scanning law & 12-240 & 140,000 & 4609 & 70 \\
    Pan-STARRS PS1 & $\sim$30000 & $grizY$ & $r$$\le$21.5 &2 same-band obs. sep. by 25 min& $\sim$67 & 35,000 & 1021 & 79 \\
    Catalina Surveys & $\sim$9000 & unfiltered & $V$$\le$19-20 & 4 obs. within 30 min &$\gtrsim$200& 32,775 & 1463 & 81 \\
    ATLAS &$\sim$13000& $c,o$ & $r$$\sim$20& 4$\times$ per night & $\sim$200 &  21061 & 484 & 81 \\
    DES single epoch & $\sim$5000 & $grizY$ & $g$$\sim$23.5 
    & irregular &$\sim50^{*}$& 5783 & 5783 & -- \\
	\enddata
    \tablecomments{The details of DES are listed for comparison. (*): by the end of the survey (Y6)}
	\label{tab:external}
\end{deluxetable*}

\citet{clementini2018} found over $1.4\times 10^5$ RRL in {\it Gaia} DR2, including $\sim 5\times 10^4$ that were previously unknown. These variables were identified from multiband ($G,G_{BP},G_{RP}$) light curves that had at least 12 observations in \textit{G} \citep[see Figure 10 in][]{holl2018}. While the {\it Gaia} temporal coverage is very uneven, their RRL catalog spans the entire sky \citep[see Figure 26 in][]{clementini2018} and has high purity $(\sim 91\%)$, making it an excellent independent check of our method at brighter magnitudes. 4609 of the {\it Gaia} DR2 RRabs were present in our initial stellar catalog (\S\ref{sec:selection}) and 3227 ($\sim$70\%) were identified as such. To assess this recovery another way, if we create a purity vs. completeness curve from these cross-matches like the one shown in Figure \ref{fig:roc}, we find an AUC of 0.727. As we have significantly fewer single-band observations than {\it Gaia} DR2, it is not surprising that we do not recover all of their RRab. 

We also searched for RRab discovered in Pan-STARRS PS1 by \citet{sesar2017}. Like DES, Pan-STARRS has sparsely-sampled multiband light curves and \citet{sesar2017} employed a similar template fitting method to identify these variables. However, \citet{sesar2017} used the final data release of PS1 with an average of 67 observations per object (compared to our median of 18). We adopted their suggested \textit{ab\_score} cut of 0.8 to select only RRab. As Pan-STARRS primarily surveyed the Northern hemisphere, we found just 1021 RRab in our initial stellar catalog, but we identified $805$ ($\sim$79\%) as such, with an AUC of 0.681. As the Pan-STARRS light curves are the most similar to the DES Y3 ones out of all the external catalogs under consideration, our similar classification results show that our approach is similarly effective as the methods used by \citet{sesar2017}.

The Catalina Surveys RRL catalog \citep{drake2013,drake2013stream,drake2014,torrealba2015catalina,drake2017} is based on a wide-field (26,000 deg$^{2}$) time series survey that probes the variable sky to a depth of $V\sim 19-20$~mag. The observations are unfiltered and collected in sequences of four images equally spaced over 30 minutes in each pointing \citep{drake2009}. After several years of operation, the Catalina Surveys have over 200 observations for most of their variables \citep{drake2014}, which makes the catalog largely complete. Given the limited magnitude overlap between the Catalina Surveys and DES, we only found 1463 of their 32775 RRab in our initial stellar catalog, but we identified 1185 ($\sim$81\%) as such, with an AUC of 0.733. 

ATLAS, a planetary defense initiative with a high cadence well suited for variability studies, recently released its first catalog of variable stars \citep{heinze2018}. Thus far, ATLAS has at least 200 observations across two filters ($c,o$) over one-fourth of the sky. We select RRab stars from the ATLAS DR1 variable star catalog using the suggested CasJobs query in Appendix 10.2 of \citet{heinze2018}. As ATLAS is based in the Northern hemisphere and quite shallow  compared to DES ($r\approx 20$~mag), we only have 484 of their 21061 RRab in our initial stellar catalog but identify 391 ($\sim$ 81\%) as such, with an AUC of 0.635. This recovery rate is quite similar to the ones for Pan-STARRS and the Catalina Surveys.

In addition to searching for RRab candidates with previous identifications from the aforementioned wide-field surveys, we also checked for overlaps near the Magellanic Clouds \citep{ogle2016}, the Fornax dSph \citep{bw2002}, the Sculptor dSph \citep{martinezvazquez2016}, in the General Catalogue of Variable Stars \citep{samus2017}, and in the SIMBAD database \citep{simbad2000}. To the best of our knowledge, and based on publicly available catalogs, 1795 (nearly 31\% of our sample) are newly-discovered RRab candidates. Although the external catalogs under consideration are not complete, the fraction of their RRab recovered by our analysis is consistent with our estimate of $\sim75$\% completeness. Our method is just as effective (if not more so) at recovering RRL from sub-optimally sampled data than the methods used in comparable surveys. 

Although we recover most of the RRab in the aforementioned overlapping catalogs, we can see from the AUC of each of these that there is a marked degradation in our algorithm's performance when applied to light curves outside our S82 training set. Thus, we use their AUC values to construct a confidence interval for the performance of our classifier. With the AUC of the training set and all four of these external cross-matches, we find a mean AUC of 0.728 with a standard deviation of 0.077. From this, we can determine that our classification methods have a lower effiency for fainter RRab. Unfortunately, we do not have well-characterized training data in a comparable filter system for fainter RRab, so we tested this with simulated light curves.

\subsection{Estimated Recovery Rates and Uncertainties from Simulated Data}\label{sec:simresults}

\begin{figure*}
	\includegraphics[width=1\textwidth]{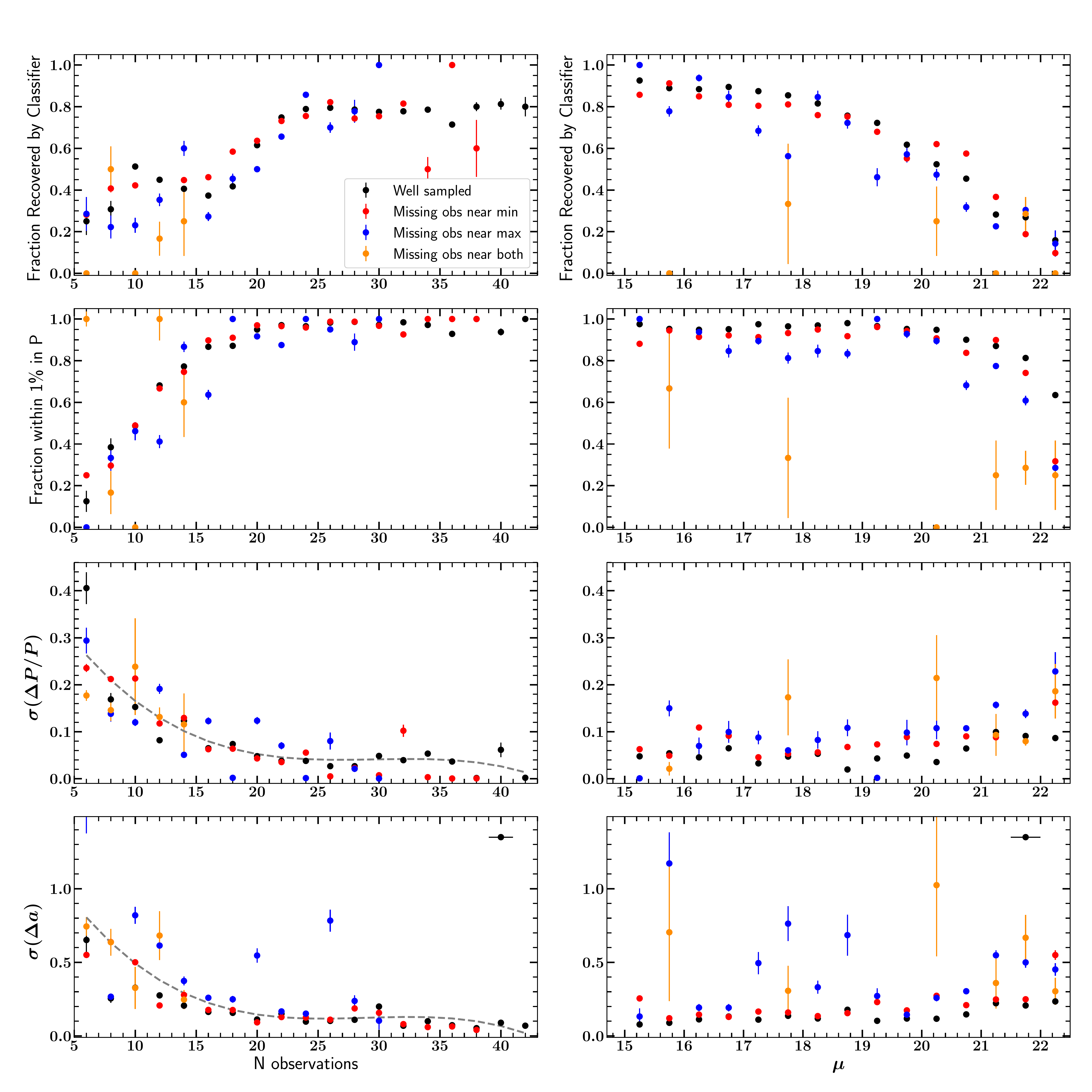}
	\caption{Recovery rates and parameter uncertainties as a function of the number of observations in the light curves and distance modulus $\mu$. Colored points denote the behavior of the recovery fractions or parameter offsets for light curves with the phase sampling flags described in \S\ref{sec:visual_validation}. Uncertainties on these values were estimated using jackknife resampling. The black points in the upper right corner of the bottom panels show the representative width of the bins in each column. Dashed grey lines show the best fitting 3rd degree polynomial to the trends shown by the combined simulated data used to assign uncertainties in the RRab catalog. The coefficients for these fits are listed in Table \ref{tab:param_uncertainties}. \textit{First row:} Fraction of simulated RRab light curves which received a classifier score $\ge 0.35$. \textit{Second row:} Fraction of estimated periods within 1\% of the true input values. \textit{Third row:} Standard deviation of the percent difference in period. \textit{Bottom row:} Standard deviation of the offset in amplitude. \textit{Note:} If not visible, uncertainties are smaller than the plotting symbols.}
    \label{fig:param_uncertainties}
\end{figure*}

To estimate the robustness of our results for the noisier photometry at fainter magnitudes, we followed a method similar to \citet{medina2018} and applied our method to simulated light curves with known light curve parameters in the DES filter system. We created the simulated light curves by sampling the smoothed templates of \citet{sesar2010} in {\tt gatspy} \citep{vanderplas2015} with the DES cadence from different areas of the survey. We shifted these light curves to various distances by adding the appropriate distance modulus and inserting scatter in the observations based on the magnitude-dependent uncertainty relations we found in \S\ref{sec:rescale} (shown in Figure \ref{fig:rescale_errors}). Appendix \ref{sec:simulated} contains further details on the construction of these simulated light curves. 

Figure \ref{fig:param_uncertainties} shows the recovery rates of both the classifier and the period as a function of magnitude and total number of observations. As expected, the recovery rate of our algorithm decreases significantly with increasing distance modulus. This is mostly due to the larger photometric uncertainties and fewer observations due to the brighter limiting magnitudes for the redder bands (see \S\ref{sec:y3q2}). We see that the accuracy of the period estimation decreases following the trend of increasing photometric errors shown in Figure \ref{fig:param_uncertainties}, and dramatically improves with increasing total number of observations up to $N\sim20$. As expected, the RRab classification accuracy follows a similar trend. We find that our template fitting recovers the true period to within 1\% for 95\% of the simulated light curves with N=20 observations.

Beyond assessing our classifier performance with these simulated light curves, we can also use them to estimate the uncertainties of the best fitting template parameters. To make sure we treat light curves with especially poor phase coverage separately, we divided the simulated light curves into groups based on their ``flag\_minmax" values (described in \S\ref{sec:visual_validation}). Then, we subdivided those into bins of two observations and 0.5 magnitude wide in N and $\mu$, respectively. In each of these bins, we calculate the fraction of light curves with period estimates within 1\% of their input values for each phase sampling group. To quantify the uncertainty of the period estimates, we calculated the standard deviation of $\Delta P/P = P_{est}- P_{true}/P_{true}$, where the ``est" subscript represents the parameter estimate from the template fitting and ``true" represents the input value of the simulated light curve. Likewise, we calculated $\Delta a = a_{est} - a_{true}$ to quantify the uncertainty of the amplitude estimates. The number of light curves included in each bin differs widely, so we estimate the spread of these uncertainty values within each subgroup with jackknife resampling. These results are shown in Figure \ref{fig:param_uncertainties}.

Other than fluctuations due to the small sample sizes in some of the bins, these values follow expected trends. When there are fewer observations to constrain the parameter values during the template fitting, both the period and the amplitude are more uncertain, with these values beginning to stabilize around N=20 observations. In distance space, the parameter estimates are generally low until $\mu \approx 20$, where the brighter detection limits of the redder filters decrease the number of observations in the light curves. We have very few simulated light curves that are missing observations near their maximum only or both their maximum and minimum (the blue and orange points in Figure \ref{fig:param_uncertainties}), so we cannot draw any definitive conclusions about the effect of phase sampling on the estimation of these parameters. We assign these parameter uncertainties to the real RRab candidates based on the best fitting 3rd degree polynomial to the trends in N observations for all simulated light curves. We do not assign uncertainties to objects with $N > 43$ observations due to a lack of simulated data with that sampling. We also do not report these uncertainties for objects not identified as RRab by the classifier since these simulated light curves do not accurately represent the behavior of non-RRab. The coefficients of the best fitting polynomials are included in Table \ref{tab:param_uncertainties} and the uncertainties are included in the full catalog described in Appendix \ref{sec:appendix_data_products}.

\begin{deluxetable}{lcllr}
	\tabletypesize{\footnotesize}
    
	\tablecaption{Coefficients for Parameter Uncertainties}
    \tablehead{
    \colhead{Value} & \colhead{$p_{0}$} & \colhead{$p_{1}$}&\colhead{$p_{2}$}&\colhead{$p_{3}$}}
	\startdata
    $\sigma(\Delta P /P)$ & 4.8585$\times 10^{-1}$ & -4.5912$\times 10^{-2}$ & 1.5636$\times 10^{-3}$ & -1.7574$\times 10^{-5}$\\
    $\sigma(\Delta a)$    & 1.5333 & -1.5101$\times 10^{-1}$ & 5.3006$\times 10^{-3}$ & -6.1034$\times 10^{-5}$\\
	\enddata
    \tablecomments{The best fit 3rd degree polynomial is of the form $\sigma(\mathrm{Value}) = p_{0} + p_{1}N + p_{2}N^{2} + p_{3}N^{3}$.}
    \label{tab:param_uncertainties}
\end{deluxetable}

The uncertainty of the remaining parameter $\phi$ is significantly more difficult to constrain. Phases for individual observations in the folded light curves are calculated using $\mathrm{phase} = (\mathrm{MJD}/P)\mod 1$. Any small offset in the period will compound over successive pulsations and result in a phase offset that varies over time.  Even simulated light curves with $\Delta P /P < 0.0005$ (a difference $<1$ minute) can yield $\Delta \phi \approx 0.5$ after three years when compared to the phases calculated using the input period. Thus, we do not report these uncertainties in $\phi$ as they  require a level of period precision we do not attain even in light curves with $N>20$. We caution against using the phases reported here for purposes other than plotting the template curves.

\subsection{Uncertainties in the Distance Moduli}\label{sec:uncertainties}

\begin{figure}[htb]
  \includegraphics[width=0.5\textwidth]{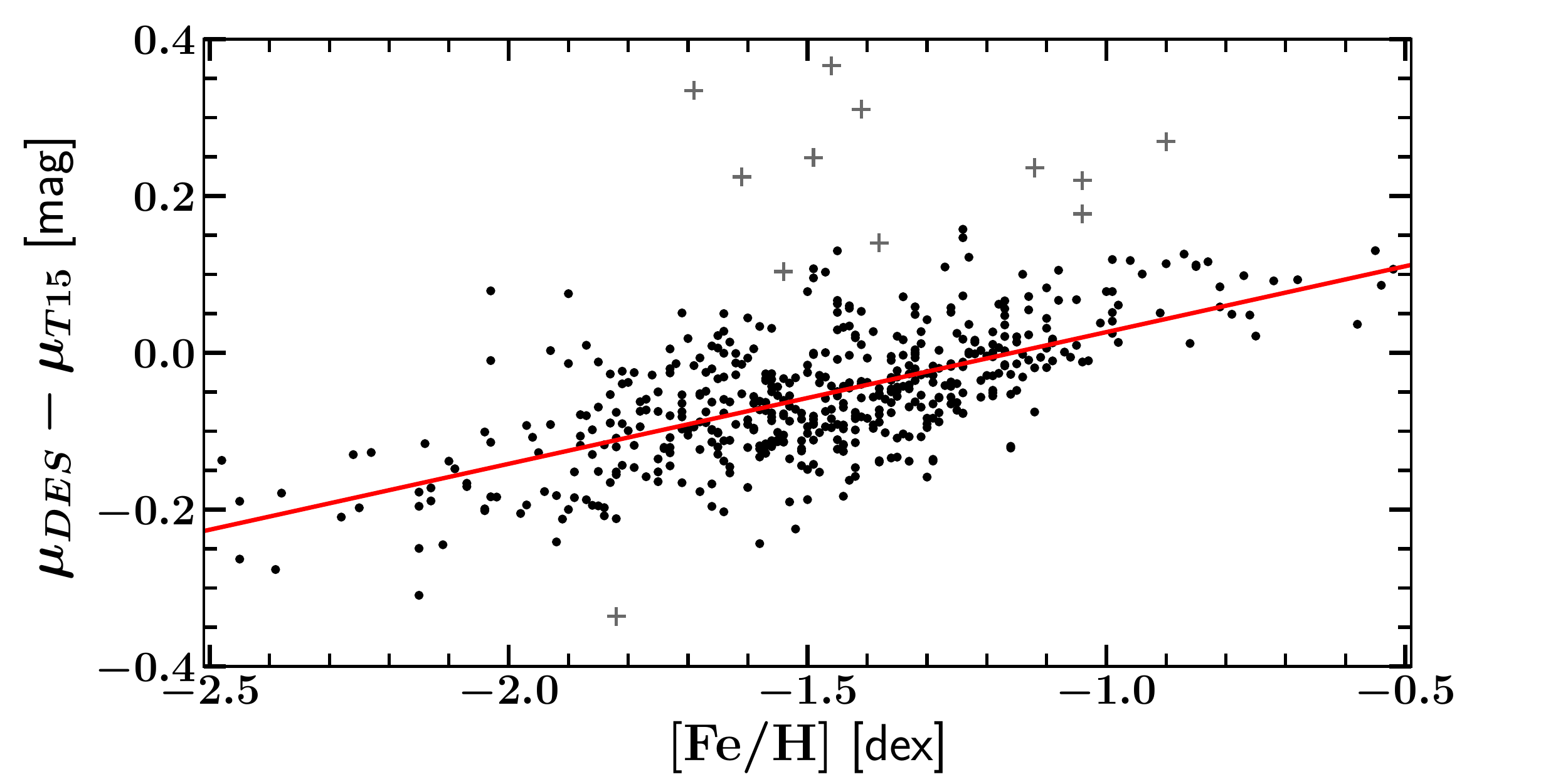}
  \caption{Difference in distance moduli between our work and \citet[][T15]{torrealba2015catalina} plotted against their photometric metallicity estimates. Stars that passed the 3-$\sigma$ clip are plotted as black points, while the points that were removed are plotted as grey crosses. The best-fit linear relation fit to the clipped data is plotted in red. The RMSE of the fit is 0.06.} 
  \label{fig:feh_uncertainties}
\end{figure}

Since the absolute magnitudes of RRL depend on their metallicities (see Figure 14 in \citealt{marconi2015}), the fixed [Fe/H] of our RRL model contributes systematic uncertainty to our distance estimates. Although the abundances of the individual RRL in our catalog are unknown, we can approximate the size and direction of this effect by comparing our results to those from an external catalog with metallicity measurements. The Catalina Surveys catalog of \citet{torrealba2015catalina} (hereafter T15) is convenient for this purpose because it has photometric estimates of [Fe/H] and a significant overlap with the DES survey footprint (although it has a brighter magnitude limit, see \S\ref{sec:externalcats}. We calculated the difference in distance moduli for 521 RRL in common between both catalogs, $\Delta\mu = \mu_{DES}-\mu_{T15}$, which we plot as a function of [Fe/H] in in Figure \ref{fig:feh_uncertainties}. 

We split the sample into bins of 0.1 dex in metallicity and perform an iterative 3-$\sigma$ clip from the median value using the {\tt sigma\_clip} function in {\tt astropy.stats}. We fit a linear relation between $\Delta\mu$ and [Fe/H]:

\begin{equation}
\begin{aligned}
\Delta\mu (\textrm{DES} - \textrm{T15}) = (-0.058 \pm 0.003) \\
+ (0.168 \pm 0.009)\ ([\textrm{Fe/H}]+1.5).
\label{eq:feh_fit}
\end{aligned}
\end{equation}

The root-mean-square error (RMSE) of the fit is $0.06$~mag, consistent with the standard deviation of $\Delta D/D$ between this work and \citet{sesar2010} listed in Table~\ref{tab:accuracy}. Thus, we estimate our statistical uncertainty in distance moduli to be $\sigma_{\mathrm{stat}} = 0.06$~mag. 

As is evident from Figure \ref{fig:feh_uncertainties}, our algorithm systematically underestimates distances for very metal-poor RRL and overestimates distances for more metal-rich RRL. While this cross-matched sample illustrates how much of an effect an RRab's metallicity has on the accuracy of our distance estimates, we caution against using this sample to derive a metallicity correction to our distance estimates. The T15 RRL cover most of the spatial DES footprint, but our RRab catalog extends to fainter magnitudes than the magnitude limit probed by the T15 sample, which means we cannot assume the metallicity distribution of this sample accurately reflects that of the entire catalog. However, if we use this subsample and assume that the stellar halo metallicity distribution function is represented by a Gaussian with a mean of $[\mathrm{Fe/H}]=-1.5$~dex and standard deviation of $\sigma=0.3$~dex \citep{ivezic2008z} as in \citet{sesar2017}, we find a 1-$\sigma$ systematic uncertainty of $\sigma_{\mathrm{[Fe/H]}}\approx 0.05$~mag in distance modulus. If we follow \citet{medina2018} and estimate distance offsets for a metallicity shift of $\pm 0.5$ dex and $\pm 1.0$ dex in this subsample, we find a change in distance modulus of $0.08$~mag and $0.17$~mag, respectively. Again, we caution that the true distribution of metallicities in our full catalog is unknown, so these values are merely representative of the systematic uncertainty that would apply to particular stellar populations in the Milky Way halo. However, given our lack of metallicity information, we cannot quantify these systematic offsets without making such assumptions.

Another contribution to the systematic uncertainty we have not previously considered is the RRL evolution off the horizontal branch. We adopt the value $\sigma_{<V>} = 0.08$~mag, which \citet{vivas2006} estimated from RRL in globular clusters (see their \S4 and Figure 4).  Adopting the halo metallicity distribution from \citet{sesar2017}, we add both sources of uncertainty in quadrature to arrive at $\sigma_{\mathrm{sys}} = (\sigma_{\mathrm{[Fe/H]}}^{2} + \sigma_{<V>}^{2})^{1/2} \approx 0.09$~mag.

We verify this estimate of systematic uncertainty by comparing our estimated distance moduli for various MW satellites with previously-published results. For the Fornax dSph, which has a horizontal branch [Fe/H]$\approx -$1.8~dex \citep{rizzi2007}, our median distance modulus is $0.05\pm0.04$~mag closer than the values of $\mu=20.72 \pm0.04$ and $\mu=20.72 \pm0.06$~mag found by \citet{greco2006} and \citet{rizzi2007}, respectively. In the case of the Sculptor dSph, we find a median distance modulus $0.13\pm0.04$~mag closer than the value of $\mu=19.62\pm0.04$~mag from \citet{martinezvazquez2016}. This larger difference is likely due to the large spread in metallicity exhibited by Sculptor's stellar populations, $-2.3 \lesssim [Fe/H] \lesssim -1.5$~dex \citep{martinezvazquez2016}. Our distance estimate to the LMC, based on the RRL we found in its outskirts, is $0.12\pm0.09$~mag closer than the $\mu=18.52\pm0.09$~mag found by \citet{clementini2003}. The LMC also has a spread of metallicities for HB stars, centered on [Fe/H]$\approx-$1.5~dex with a dispersion of 0.4~dex \citep{clementini2003,gratton2003}. We expect that replacing the template's $M_{b}(P)$ relation with a calibrated \textit{P-L-Z} or \textit{P-L-C} relation in the DECam filters (K.~Vivas et al., in prep.) will significantly reduce these offsets.

In summary, our distance moduli have 1-$\sigma$ statistical and systematic uncertainties of 0.06 and 0.09 mag, respectively. The equivalent distance uncertainties are $\sim$2.8\% (stat) and $\sim$4.2\% (sys).  


\section{Discussion} \label{sec:discussion}

\subsection{Detection Biases}\label{sec:detection_biases}

\begin{figure}
	\includegraphics[width=0.5\textwidth]{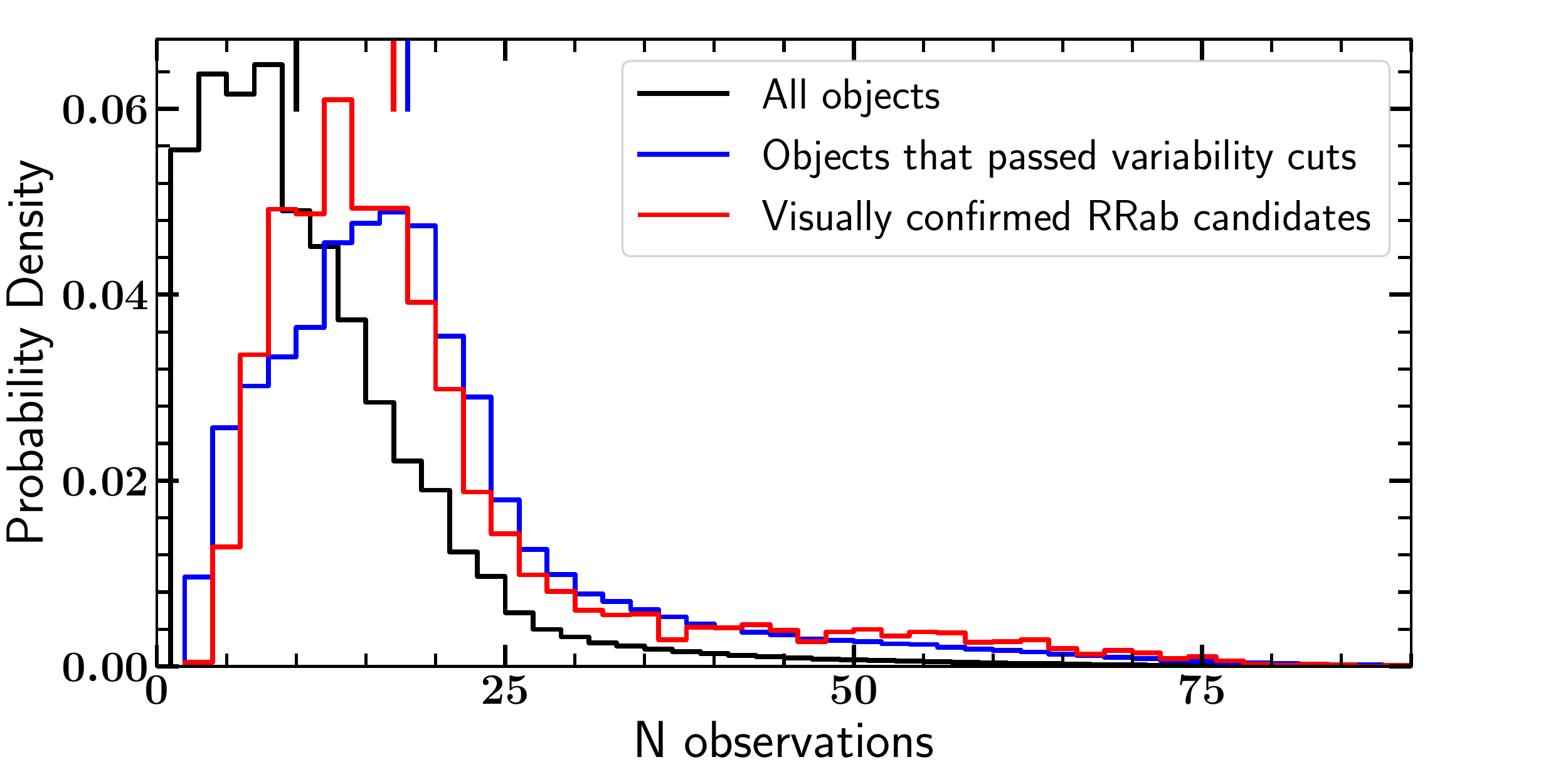}
	\caption{Histograms of the total number of observations across all bands for objects in the stellar sample (see \S\ref{sec:selection}) (black), objects which passed the variability cuts (blue), and the objects identified as RRab (red). Short line segments denote the median number of observations for each group (10, 18, and 17, respectively). Note that many of the RRab candidates have very few observations and would benefit from follow-up observations to confirm their nature.}
    \label{fig:nhist}
\end{figure}

The strength of DES lies in its wide-field coverage and depth, but the results presented here are limited by the low number of multiband observations. Figure \ref{fig:nhist} displays histograms of the total number of observations for all objects in the stellar sample (black), all objects passing variability cuts (blue), and all RRab candidates (red). The median number of total observations for each group, marked by a short colored line segment, is 10, 18, and 17, respectively. Note that most of our RRab have fewer observations than the N$\sim$20 observation threshold we saw from the simulation results in Figure \ref{fig:param_uncertainties}. As future DES data releases will have an increased number of observations, we expect to find more RRL and have a more robust classification of the candidates presented here.

We note that the light curves used in this analysis typically had fewer total observations than the number expected from three years of DES data. We suspect that the total number of observations for the ``stellar" sample is skewed by objects near the detection limits of DES, which suffer from noisy photometry and likely have few overlapping observations across all five filters. This low number of observations is also a result of the stringent quality cuts we applied on the single epoch photometry in \S\ref{sec:selection}. In future work, we aim to be more judicious in applying our photometric quality cuts so that we do not discard observations unnecessarily.  

\begin{figure*}[!htb]
	\includegraphics[width=1\textwidth]{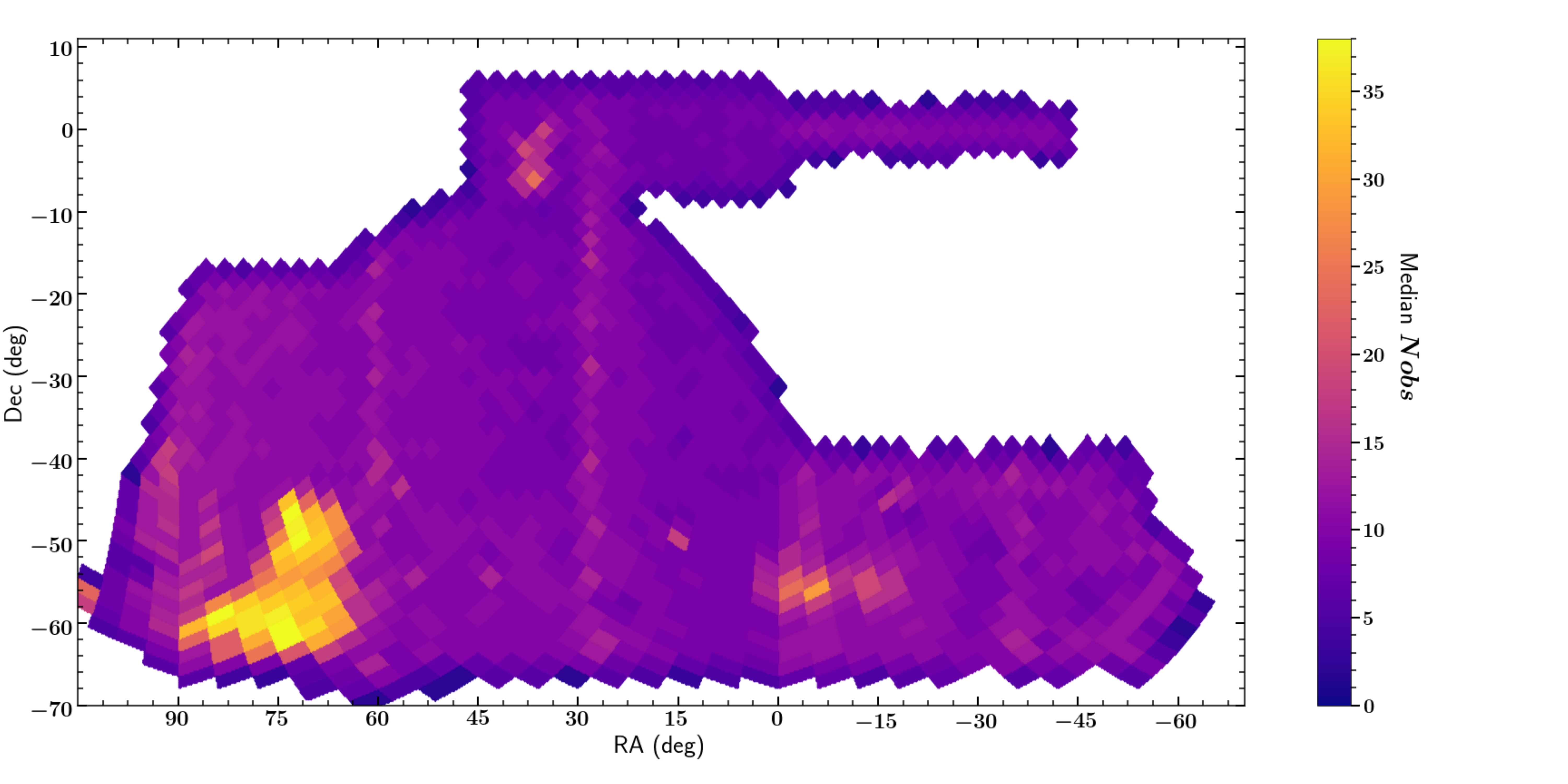}
	\caption{Spatial map of the median number of observations across all bands in each HEALPix. As expected, the regions with the fewest observations are near the edges of the survey region. The regions with a relatively large number of observations correspond to Science Verification or supernova fields. Outside these regions, the DES coverage is relatively uniform, but suffers from a small number of time series observations. Future studies of RRL will benefit from additional years of DES data.}
    \label{fig:nspatial}
\end{figure*}

To verify that our sample is not affected by spatial fluctuations in the number of observations, we calculated the median number of total observations in each HEALPix of our Y3Q2 stellar data set. We show the median and the standard deviation of the total number of observations of light curves in each HEALPix in Figure \ref{fig:nspatial}. As expected, regions with the lowest number of observations fall near the edges of the survey footprint. Regions which have a median number of observations $\ge25$ correspond to the Science Verification region, in which 50 observations were made in the first year to demonstrate year 5 depth, and the DES Supernova fields, which are observed roughly weekly \citep[e.g.][]{des2016}. The linear patterns of constant Right Ascension are a result of the survey observation strategy \citep{abbott2018}. Beyond these patterns, the DES photometry suffers in photometric completeness in crowded stellar fields near the central regions of nearby dSph galaxies and globular clusters, thus our catalog also suffers in completeness near those regions. Otherwise, the survey coverage is fairly uniform and we do not expect large scale trends in RRL detection outside of these fields of larger-than-average observation counts and dense stellar populations. We expect the addition of DES Y4-Y6 data to increase our detections of RRL considerably.

Some additional biases in our RRab sample are results of choices made to exclude non-RRab from our analysis. While we weighted our initial variability cuts by the photometric errors to make the cuts robust against spurious observations (see \S\ref{sec:varcuts}), using these error-weighted metrics biased our variable sample against RRab with smaller amplitudes located at larger distances. We also excluded some real RRab from our sample by limiting the period range to $0.44$~d $\le P \le 0.89$~d to avoid the common 1-day alias.

\subsection{Spatial Distribution of the Candidates}\label{sec:spatial_discussion}

\begin{figure*}[!h]
\centering
\includegraphics[width=1\textwidth]{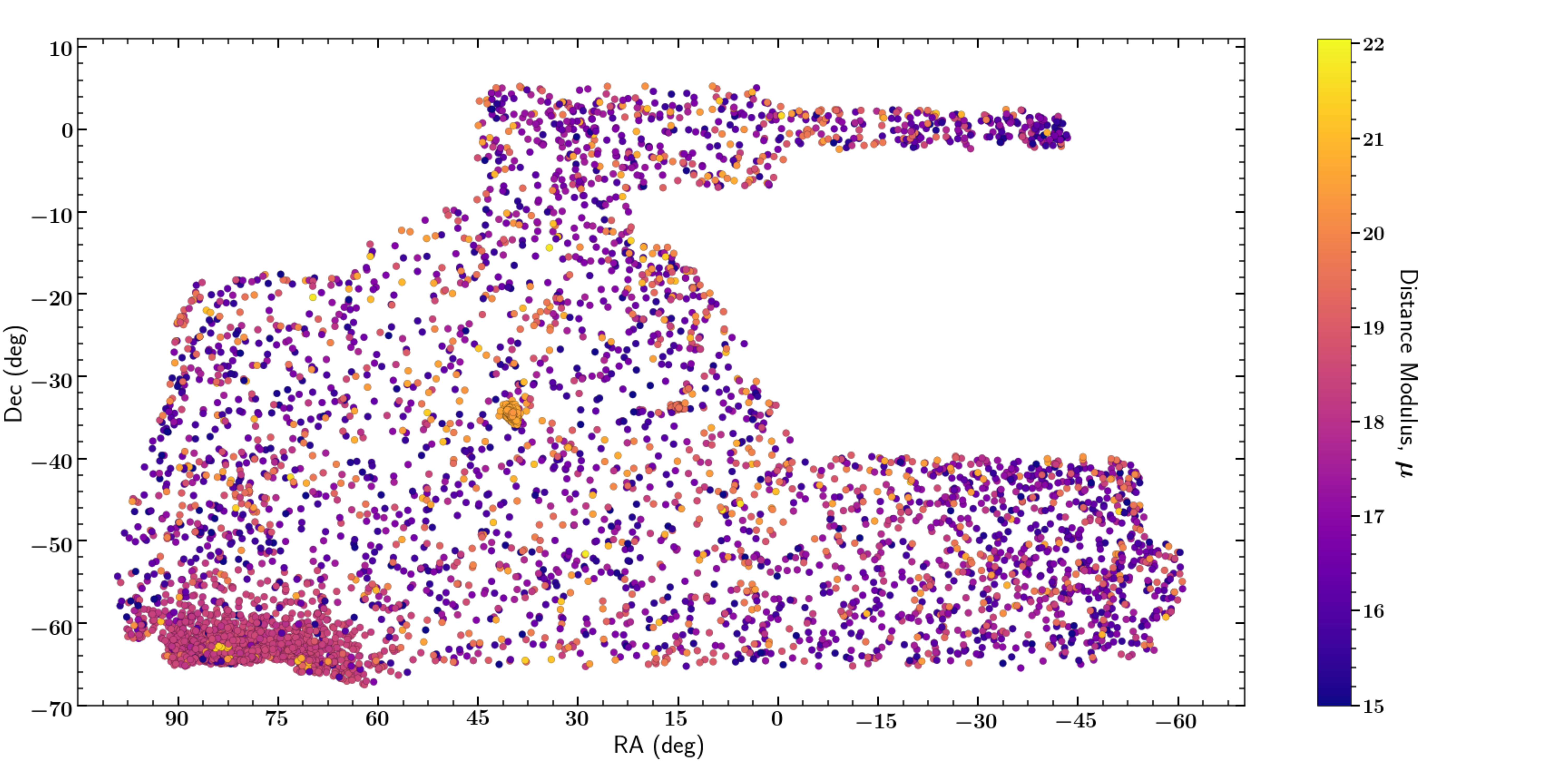}
\caption{Map of 5783 visually accepted RRab candidates across the DES wide-field survey footprint. The RRab are marked by dots colored by distance modulus. Large MW satellite galaxies are easily distinguishable by their overdensities of RRab. The outskirts of the LMC are located near $(80\degree,-62\degree)$, the Fornax dSph is located near $(41\degree,-34\degree)$, and the Sculptor dSph is located near $(15\degree,-34\degree)$.} 
\label{fig:spatial}
\end{figure*}

\begin{figure*}[!hbt]
  \includegraphics[width=1\textwidth]{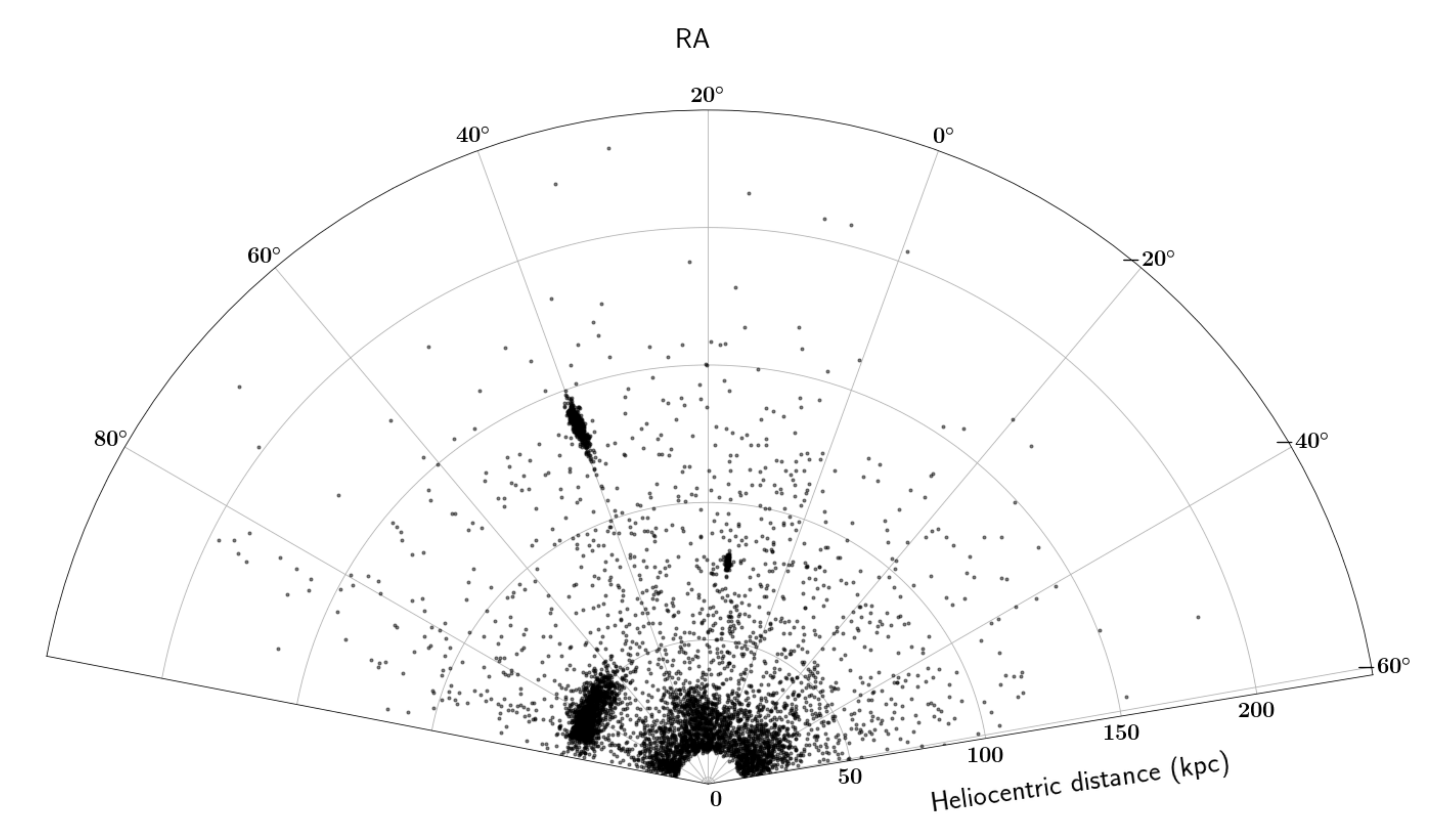}
  \caption{Radial Distribution of DES RRab stars. The overdensities associated with (in order of decreasing heliocentric distance) the Fornax dSph, the Sculptor dSph, and the periphery of the Large Magellanic Cloud are easily distinguishable. Note: The Sculptor and Fornax galaxies appear elongated due to uncertainties in the RRab distance moduli.}
  \label{fig:radial}
\end{figure*}

\begin{figure*}[!hbt]
  \includegraphics[width=1\textwidth]{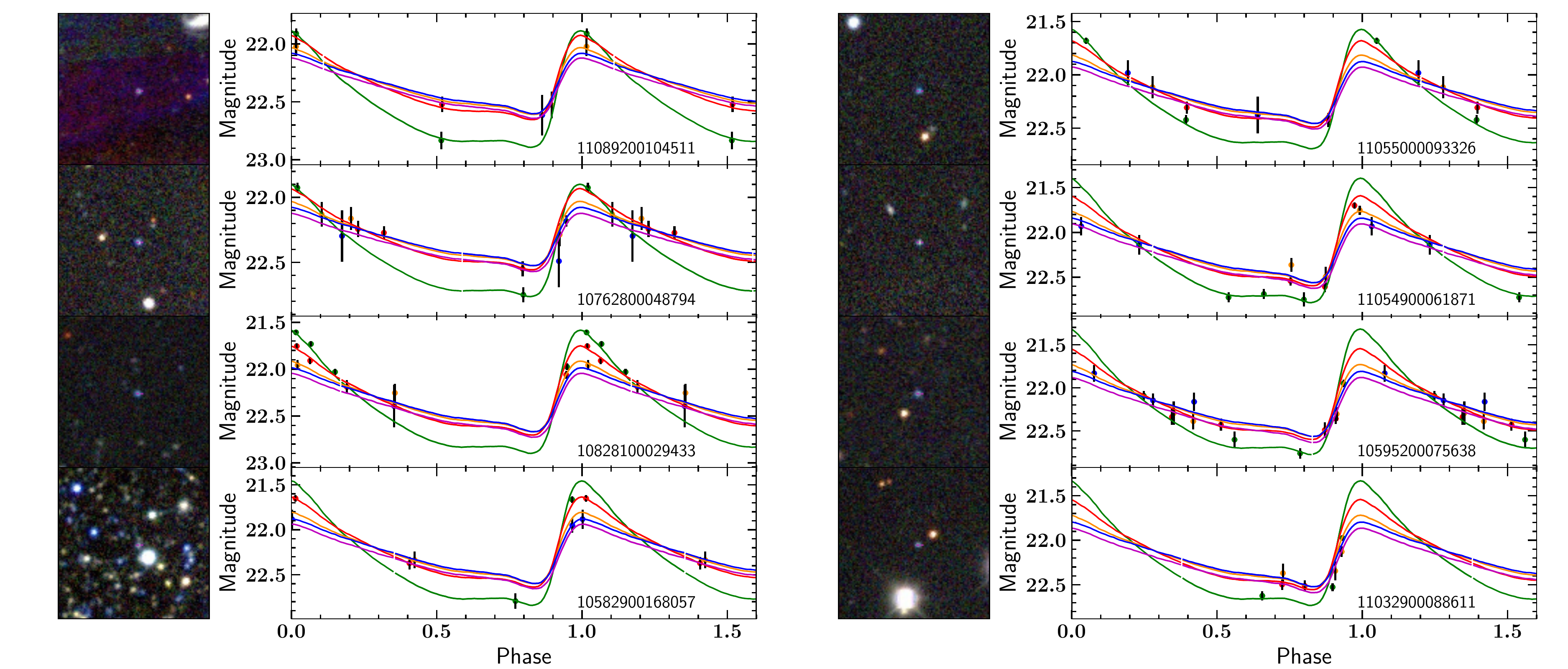}
  \caption{DES coadded images and representative light curves of visually-accepted candidates beyond 200 kpc, labeled with their Y3Q2 ID number. The observations and templates are colored by filter using the same convention as Figure \ref{fig:template}. These very distant candidates have very few observations and will benefit from future DES data releases.}
  \label{fig:d>200kpc}
\end{figure*}

The spatial distribution of the 5783 visually validated RRab candidates is shown in Figure \ref{fig:spatial}. We also plot these candidates as a function of their heliocentric distance in Figure \ref{fig:radial}. In both figures, the overdensities of RRab candidates associated with (in order of decreasing heliocentric distance): the Fornax dSph, the Sculptor dSph, and the the outskirts of the Large Magellanic Cloud are easily visible. We expect that the inclusion of a metallicity term into the model combined with additional epochs of DES observations in the next release of this catalog will enable further characterization of these and other substructures.

One of the largest strengths of the DES data set is its depth (see our comparisons to other wide-field surveys in Figure \ref{fig:externalcats}). This is extremely valuable for our understanding of the outer halo as the current census of RRL known beyond 100 kpc falls short of the thousands predicted by simulations \citep{sanderson2017}. In this work, we identified 800 RRab candidates beyond 100 kpc (most of which have been previously discovered) and eight RRab candidates beyond 200 kpc, all of which are new discoveries. The coadded images and light curves for the candidates beyond 200 kpc are shown in Figure \ref{fig:d>200kpc}.

The three most distant visually verified RRab candidates in our sample have heliocentric distances of $\sim231.6$, 223.0, and 221.3 kpc. While these three stars are the most distant to-date RRab in the Milky Way, they are not the most distant RRL. \citet{medina2018} recently found two RRc with larger distances (232.9 and 261.2 kpc) using data from the HiTS Survey \citep{forster2016}. Even though the candidates in our RRab sample suffer from a small number of observations and require additional follow-up for confirmation, the fact that there are so many RRab beyond 100 kpc and three RRab beyond 220 kpc provide reasonable evidence that the Milky Way stellar halo extends at least out to 220 kpc. Future DES data releases and other upcoming deep surveys such as LSST will increase the census of known RRL at this distance, enabling further characterization of the outer halo.

\subsection{Applicability to LSST}
\label{sec:lsst}

The next-generation large ground-based Large Synoptic Survey Telescope (LSST) \citep{ivezic2008lsst} is set to begin full science operations in early 2023\footnote{\url{https://www.lsst.org/about/timeline}}. The most current LSST ``Baseline Cadence" for its Wide-Fast-Deep Survey (WFD), which covers 18,000 deg$^{2}$ of the sky and comprises $\sim$85\% of its total allocated observing time, is to image each field twice 40 minutes apart once every three days in a different filter. After 10 years of operation, each field is expected to have a median of (62, 88, 199, 201, 180, 180) visits in $ugrizy$ with a single epoch depth of (23.14, 24.47, 24.16, 23.40, 22.23, 21.57) mag. Assuming these observations are spaced uniformly over 10 years, one can expect most light curves to have $\sim$80 multiband observations within the first year \citep{lsst2017}\footnote{See the most current version of the draft of the LSST Observing Strategy white paper located at \url{https://github.com/LSSTScienceCollaborations/ObservingStrategy}.}.

\citet{oluseyi2012} found from their analysis of simulated LSST data that reliable RRL period estimation will require several years of operation, however, several multiband techniques have been developed since their publication. For instance, \citet{vanderplas2015} estimated accurate periods 64\% of the time on downsampled S10 light curves with $\sim$55 observations and \citet{sesar2017} accurately estimated periods for 85\% of their PS1 training set with $\sim$67 observations. Our simulations show that our template fitting method is capable of estimating the correct periods to within 1\% for 95\% of the light curves with 20 total observations. Thus, our algorithm would be effective to identify potential RRab candidates (which would  need followup for confirmation) within the first year of LSST operations. After the first year, the light curves in the WFD survey will be adequately sampled to use other multiband methods available in the literature. 

\section{Summary} \label{sec:highlight}

We have presented a new physically-motivated general multiband RRab template and a computationally-efficient fitting procedure. We combined this method with a random forest classifier to create a powerful technique that can robustly identify these variables even when fewer than 20 observations are available. Despite the poor cadence and sampling of DES data, we detected 5783 RRab candidates, 1795 (31\%) of which are previously undiscovered to the best of our knowledge. The large quantity of RRL we recovered in common with overlapping external surveys such as {\it Gaia} DR2, Pan-STARRS, Catalina Surveys, and ATLAS provide strong evidence of the effectiveness of this algorithm. Although the number of observations is relatively uniform across the survey footprint in the DES Year 3 data, time series analyses like these will benefit immensely from the additional observations in future data releases. We make the template, these catalogs, and the light curves of the RRab candidates and the training sample available to the scientific community for future studies. Our method is especially useful for other multiband data sets which were not specifically designed for time series analysis.


\acknowledgments

The authors would like to thank the anonymous referees for their insightful and constructive  comments. The authors acknowledge the Texas A\&M University Brazos HPC cluster that contributed to the research reported here: \url{brazos.tamu.edu}. KMS would like to thank Chris Stringer and Ryan Oelkers for helpful conversations. Part of this work was completed while coauthors ZL and JPL were research fellows in the Astrostatistics program at the Statistics and Applied Mathematical Sciences Center (SAMSI) in Fall 2016. ZL and JPL gratefully acknowledge SAMSI’s support. C. Nielsen and J. L.~Myron would like to acknowledge support from the NSF grant AST-1263034, ``REU Site: Astronomical Research and Instrumentation at Texas A\&M University." This research has made use of the SIMBAD database, operated at CDS, Strasbourg, France.

Funding for the DES Projects has been provided by the U.S. Department of Energy, the U.S. National Science Foundation, the Ministry of Science and Education of Spain, the Science and Technology Facilities Council of the United Kingdom, the Higher Education Funding Council for England, the National Center for Supercomputing Applications at the University of Illinois at Urbana-Champaign, the Kavli Institute of Cosmological Physics at the University of Chicago, the Center for Cosmology and Astro-Particle Physics at the Ohio State University, the Mitchell Institute for Fundamental Physics and Astronomy at Texas A\&M University, Financiadora de Estudos e Projetos, Funda{\c c}{\~a}o Carlos Chagas Filho de Amparo {\`a} Pesquisa do Estado do Rio de Janeiro, Conselho Nacional de Desenvolvimento Cient{\'i}fico e Tecnol{\'o}gico and the Minist{\'e}rio da Ci{\^e}ncia, Tecnologia e Inova{\c c}{\~a}o, the Deutsche Forschungsgemeinschaft and the Collaborating Institutions in the Dark Energy Survey. 

The Collaborating Institutions are Argonne National Laboratory, the University of California at Santa Cruz, the University of Cambridge, Centro de Investigaciones Energ{\'e}ticas, Medioambientales y Tecnol{\'o}gicas-Madrid, the University of Chicago, University College London, the DES-Brazil Consortium, the University of Edinburgh, the Eidgen{\"o}ssische Technische Hochschule (ETH) Z{\"u}rich, Fermi National Accelerator Laboratory, the University of Illinois at Urbana-Champaign, the Institut de Ci{\`e}ncies de l'Espai (IEEC/CSIC), the Institut de F{\'i}sica d'Altes Energies, Lawrence Berkeley National Laboratory, the Ludwig-Maximilians Universit{\"a}t M{\"u}nchen and the associated Excellence Cluster Universe, the University of Michigan, the National Optical Astronomy Observatory, the University of Nottingham, The Ohio State University, the University of Pennsylvania, the University of Portsmouth, SLAC National Accelerator Laboratory, Stanford University, the University of Sussex, Texas A\&M University, and the OzDES Membership Consortium.

Based in part on observations at Cerro Tololo Inter-American Observatory, National Optical Astronomy Observatory, which is operated by the Association of Universities for Research in Astronomy (AURA) under a cooperative agreement with the National Science Foundation.

The DES data management system is supported by the National Science Foundation under Grant Numbers AST-1138766 and AST-1536171. The DES participants from Spanish institutions are partially supported by MINECO under grants AYA2015-71825, ESP2015-66861, FPA2015-68048, SEV-2016-0588, SEV-2016-0597, and MDM-2015-0509, some of which include ERDF funds from the European Union. IFAE is partially funded by the CERCA program of the Generalitat de Catalunya. Research leading to these results has received funding from the European Research Council under the European Union's Seventh Framework Program (FP7/2007-2013) including ERC grant agreements 240672, 291329, and 306478. We  acknowledge support from the Australian Research Council Centre of Excellence for All-sky Astrophysics (CAASTRO), through project number CE110001020, and the Brazilian Instituto Nacional de Ci\^encia e Tecnologia (INCT) e-Universe (CNPq grant 465376/2014-2).

This manuscript has been authored by Fermi Research Alliance, LLC under Contract No. DE-AC02-07CH11359 with the U.S. Department of Energy, Office of Science, Office of High Energy Physics. The United States Government retains and the publisher, by accepting the article for publication, acknowledges that the United States Government retains a non-exclusive, paid-up, irrevocable, world-wide license to publish or reproduce the published form of this manuscript, or allow others to do so, for United States Government purposes.

\vspace{5mm}
\facilities{CTIO: Blanco (DECam) \citep{decam2015}}

\software{astropy \citep{astropy2018}, easyaccess \citep{easyaccess2018}, gatspy \citep{vanderplas2015}, Matplotlib \citep{matplotlib2007}, scikit-learn \citep{scikit-learn}, SIMBAD \citep{simbad2000}}

\appendix

\section{RRL Model Assumptions}
\label{sec:appendix_model_assumptions}

The form for the RRL model is 

\begin{equation}
 m_{b}(t) = \mu + M_{b}(\omega) + R_{b} E(B-V) + a \gamma_{b}(\omega t + \phi) 
\label{eq:tempeq}
\end{equation} 

where the population parameters, common to all RRL, are:

\begin{align*} M_b(\omega) &= \text{absolute magnitude in band $b$ for an RRL with frequency } \omega\\ R_b &= \text{total-to-selective extinction coefficient for band $b$}\\ \gamma_b &= \text{shape of the RRL light curve in band $b$} \end{align*}
The object specific parameters, different for each RRL, are

\begin{align*} \mu &= \text{distance modulus}\\ E(B-V) &= \text{reddening}\\ a &= \text{amplitude}\\ \omega &= \text{frequency $(1/P)$}\\ \phi &= \text{ phase} 
\end{align*}

For one RRL, the time-series photometry can be written as $\{\{t_{bi},m_{bi},\sigma_{bi}\}_{i=1}^{n_b}\}_{b=1}^B$ where $m_{bi}$ is the observed magnitude at time $t_{bi}$ in filter $b$ measured with (known) uncertainty $\sigma_{bi}$. The bands are indexed $1,\ldots,B$ instead of typical letters e.g., {\it ugriz}. The model and data are related by $$m_{bi} = m_b(t_{bi}) + \epsilon_{bi}$$ where $\epsilon_{bi} \sim N(0,\sigma_{bi}^2)$, meaning the noise parameter $\epsilon_{bi}$ can be viewed as a random normal variable with mean=$0$ and $Var(\epsilon_{bi}) = \sigma_{bi}^2$.

This model assumes all RRL share a common shape by band and that RRL are strictly singly periodic functions. These assumptions are an approximation. For example, our model does not account for the amplitude and phase modulations which vary according to an additional period caused by the Bla{\v z}hko effect \citep{blazhko1907}. Rather than construct a perfectly accurate model, the goal is to construct a model with few free parameters that provides a better approximation to RRL variation than existing methods. For example, a simple sinusoid model fit to 5 filters has a total of 16 free parameters (5 means, 5 amplitudes, 5 phases, and 1 frequency) while providing only a very rough approximation to the steep rise and slow decline in brightness observed in RRL light curves. In contrast, this model provides a significantly better approximation while fitting for 5 free parameters (or 4 if light curves are corrected for extinction prior to fitting).

\section{Determining Template Population Values}\label{sec:appendix_population_parameters}

We estimated the population parameters common to all RRL ($M_{b}(P)$,$R_{b}$, and $\gamma_{b}$) using a combination of theory and existing data sets. 

For the filter-dependent extinction coefficients $R_{b}$, we assumed a Galactic reddening value of $R=3.1$ from \citet{fitzpatrick1999dust}. 
These extinction values $R_{b}$ for both SDSS and DES filters are summarized in Table \ref{tab:extcoeffs}.  
Note that these values are only used if the templates are fit with light curves uncorrected for extinction. In general, it is better to fit with dust-corrected light curves because the model has one fewer free parameter and the uncertainty on distance is greatly reduced. In this work, we corrected the light curves for extinction prior to fitting the template (see \S\ref{sec:varcuts}).

\begin{deluxetable}{ccc}
	\tabletypesize{\footnotesize}
	\tablecaption{Extinction Coefficients \label{tab:extcoeffs}}
	\tablehead{\multicolumn{1}{l}{Band} & \multicolumn{2}{c}{$R_{b}$}\\ \multicolumn{1}{l}{\ \ $b$} & \multicolumn{1}{c}{\ \ \ \ \ \ SDSS\ \ \ \ \ \ } & \multicolumn{1}{c}{\ \ \ \ \ \ DES\ \ \ \ \ \ }}
	\decimals
	\startdata
	$u$ & 4.799 & N/A \\
	$g$ & 3.737 & 3.665 \\
    $r$ & 2.587 & 2.464 \\
    $i$ & 1.922 & 1.804 \\
    $z$ & 1.430 & 1.380 \\
    $Y$ & N/A & 1.196 \\
	\enddata
	\tablecomments{Based on \citet{fitzpatrick1999dust}. Note that these model dust coefficients differ from those listed in \\ \citet{schlegel1998} and \citet{abbott2018}.}
\end{deluxetable}

To develop the $P-L$ relation for this work, we determined the values of $M_b(P)$ for \textit{ugriz} using version 3.2 of the \texttt{BaSTI} synthetic horizontal branch generator\footnote{Available at \url{albione.oa-teramo.inaf.it/BASTI/WEB_TOOLS/HB_SYNT/}}, based on the evolutionary tracks of \citet{pietrinferni2004,pietrinferni2006}. We generated synthetic absolute magnitudes for RRL spanning $-0.48 \leq \log P \leq 0.08$ with a metallicity of $[Fe/H]=-2.85$ to use as our starting values for $M_{b}(P)$. Then, we shifted the template curve in each filter to match the magnitude offsets shown in real SDSS and DES light curves. We parametrized these empirical $M_{b}(P)$ by using a quadratic period-absolute magnitude $(P-L)$ relation at a fixed metallicity of $[\mathrm{Fe/H}]\approx -1.5$ (see \S\ref{sec:uncertainties} for an extended discussion of the template metallicity): 

\begin{equation}\label{eq:pl_appendix}
M_b(P) = c_{0b} + p_{1b}(log_{10}(P) + 0.2) + p_{2b}(log_{10}(P) + 0.2)^2
\end{equation}

\noindent where the $c$, $p_1$ and $p_2$ values for the SDSS and DES filters are shown in Table \ref{tab:sdss_des_pl}.

\begin{deluxetable}{lrrrrrr}
	\tabletypesize{\footnotesize}
	\tablecaption{$P-L$ Coefficients \label{tab:sdss_des_pl}}
	\tablehead{\colhead{Band $b$} & \colhead{$c_{\mathrm{SDSS}}$} & \colhead{$p_{1,\mathrm{SDSS}}$} & \colhead{$p_{2,\mathrm{SDSS}}$}& \colhead{$c_{\mathrm{DES}}$} & \colhead{$p_{1,\mathrm{DES}}$}
    & \colhead{$p_{2,\mathrm{DES}}$}}
	\decimals
	\startdata
	$u$ & 1.889 & -0.049 & -0.319 & ---   &  ---   & ---  \\
	$g$ & 0.767 &  0.167 & -0.595 & 0.730 & -0.020 & -0.065\\
    $r$ & 0.550 & -0.637 & -0.353 & 0.542 & -0.739 & 0.997\\
    $i$ & 0.505 & -1.065 & -0.202 & 0.522 & -1.136 & -0.057\\
    $z$ & 0.510 & -1.308 & -0.231 & 0.520 & -1.292 & -0.535\\
    $Y$ &  ---  &  ---   &   ---  & 0.558 & -1.392 & 0.657 \\
	\enddata
\end{deluxetable}

The light curve shape in filter $b$ $\gamma_b$ is a function of phase that covers one pulsation period. We use RRL found by \citet{sesar2010} to estimate $\gamma_b$ for the SDSS filters. We assume the same shapes for DES {\it griz} and assume the DES $Y$ band shape is the same as the $z$ band shape.

To infer the shape, we first ``fold" the well-sampled S82 RRL light curves from \citet{sesar2010} into phase coordinates by taking the modulus of the Modified Julian Dates of the observations with respect to the pulsation period of each individual RRL. We ``phase-align" the light curves by shifting them so that they all reach their maximum brightness at phase=0. We smooth the light curves by removing observations with photometric errors $\ge0.2$ and linearly interpolating them in equally spaced phase bins. Then, we shift all of the light curves so that the curves in each filter $ugriz$ have an average magnitude value $m_{b}=0$. We sample each light curve on a grid in phase space so that a single RRL is denoted by $X_{tb}$ for $t=(1,\ldots,T)$ and $b=(1,\ldots,B)$ where $t$ indexes the phase (the grid has $T=100$ equally spaced phases) and $b$ indexes the filter (total $B$ filters). 

Let $X_{itb}$ be the magnitude for the $i^{th}$ RRL, at phase $t$ in band $b$. Let $\gamma_b \in \mathbb{R}^T$ be the template in filter $b$. The template matrix of all five bands is thus defined as $\Gamma = (\gamma_1,\ldots,\gamma_B) \in \mathbb{R}^{T \times B}$. Let $a \in \mathbb{R}^n$ be the amplitudes for the $n$ RRL in the SDSS S82 sample. Let $X_{i\cdot\cdot} \in \mathbb{R}^{T \times B}$ represent the phase-folded, shifted, and normalized photometry described in the previous paragraph for the $i^{th}$ RRL. To determine the $\Gamma$ matrix of the template shapes, we solve the following optimization problem:

\begin{equation}\label{eq:optim}
\min_{\Gamma,a} \sum_{i=1}^n ||X_{i\cdot\cdot} - a_i\Gamma||_F^2
\end{equation}

\noindent where $||a||_2 = 1$ (the Euclidean norm) for identifiability and $F$ denotes the Frobenius norm, or the square root of the absolute squares of its elements. The resulting $\Gamma$ matrix are the template shapes in each filter. To reflect the dependence of the RRL amplitudes on the filter in which they were observed, we rescale the templates so the peak--to--peak $g$-band amplitude is $1$ and the amplitudes of the other filter shapes are fractions of the $g$-band amplitude. These shapes and population parameter values form the set of templates that we use for fitting in our analysis. 

\section{Fitting the Model} \label{sec:appendix_model_fitting}

In this section, we describe how to fit the model to the data. Directly using the inverse of the observation uncertainties as weights is known to be suboptimal when the templates are an approximation, see \citet{long2017}. We estimate a model error term $\sigma_{me}$ which is then used in the least square fitting. To compute $\sigma_{me}$, we fit the template to all-well sampled SDSS RRL light curves and compute the difference between the squared residuals and the squared photometric error $\sigma^2_{bi}$. $\sigma_{me}$ is the square root of the average of these differences. The value across all of the SDSS bands $ugriz$ is 0.0547. 

The model is fit by minimizing a weighted sum of squares (``$\chi^2$ minimization"). There are at most five free parameters $\mu$, $E[B-V]$, $a$, $\omega$, $\phi$. The dust can be turned off in the fitting in which case $E(B-V)$ is set to $0$. We perform a grid search across the frequency because the objective function is highly multimodal. At frequency $\omega$ in the grid, we solve for the four parameters $\mu, E(B-V), a, \phi$ using:

\begin{equation}\label{eq:gridest}
\min_{\mu,E(B-V),a,\phi} \sum_{b,i} \frac{(m_{bi} - M_b(\omega) - \mu - E(B-V)R_b - a\gamma_b(\omega t_{bi} + \phi))^2}{\sigma_{bi}^2 + \sigma_{me}^2}
\end{equation}

\noindent where $b$ is the filter index and $i$ is the epoch index. We use a block--relaxation method in which we alternate between minimizing across the $(\mu,E(B-V),a)$ parameters and minimizing across the $\phi$ parameter. The number of iterations can also be specified.

When minimizing across $(\mu,E(B-V),a)$
at fixed $\phi$, the model is linear in $(\mu,E[B-V],a)$, so we find the closed-form weighted least squares solution. Occasionally the update will result in a negative amplitude. In this case, we do a random phase update in the next step (i.e. draw phase uniformly in [0,1]), rather than the Gauss--Newton method described below.

When minimizing across $\phi$,
with fixed ($\mu,E[B-V],a$), we cannot analytically solve for $\phi$. Instead we use a Gauss--Newton. Define

\begin{equation}\label{eq:gaussm}
m^*_{bi} \equiv m_{bi} - M_b(\omega) - \mu - E(B-V)R_b
\end{equation}

and 

\begin{equation}\label{eq:gaussgam}
\gamma_{bi}(\phi) \equiv a\gamma_b(\omega t_{bi} + \phi)
\end{equation}

Then the objective function which we seek to minimize is 

\begin{equation}\label{eq:gaussphi}
g(\phi) = \sum_{b,i} \frac{(m^*_{bi} - \gamma_{bi}(\phi))^2}{\sigma_{bi}^2 + \sigma_{me}^2}.
\end{equation}

With $\phi^{(m)}$ as our current phase estimate, the Newton update has the form: 

\begin{equation}\label{eq:newtonupdate}
\phi^{(m+1)} = \phi^{(m)} - H(g)^{-1}(\phi^{(m)})\nabla(g)(\phi^{(m)}) 
\end{equation}

where $\nabla(g)$ and $H(g)$ are the first and second derivatives of $g$. We have 

\begin{equation}\label{eq:fderivg}
\nabla(g) = \frac{\partial g}{\partial \phi} = - 2\sum_{b,i} \frac{(m_{bi}^* - \gamma_{bi}(\phi))\gamma_{bi}'(\phi)}{\sigma_{bi}^2 + \sigma_{me}^2}. 
\end{equation}

and 

\begin{equation}\label{eq:sderivg}
H(g) = \frac{\partial^2 g}{\partial \phi^2} = 2 \sum_{b,i} \frac{\left(\gamma'_{bi}(\phi)^2 - (m_{ib}^* - \gamma_{bi}(\phi))\gamma''_{bi}(\phi)\right)}{\sigma_{bi}^2 + \sigma_{me}^2}. 
\end{equation}

The Gauss--Newton update approximates $H(g)$ with 

\begin{equation}\label{eq:gaussnewton}
H^*(g) = 2 \sum_{b,i} \gamma'_{bi}(\phi)^2.
\end{equation}

\noindent where we substitute $H^{*}$ for $H$ in Equation \ref{eq:newtonupdate}, rather than using $H(g)$ in Equation \ref{eq:sderivg}. This is a standard approach in non--linear regression which avoids computation of $\gamma''_{bi}$ and ensures that the second derivative is positive (see Section 14.4 in \citet{lange2010}). We approximate $\gamma'_{bi}$ by storing numerical derivatives of the $\gamma_b$ templates. At each new $\omega$ in the grid of frequency we obtain a warm start for the $\mu,E[B-V],a,\phi$ parameters by using estimates from the last frequency. We choose the frequency at which the RSS is minimized to be the parameters of the best-fitting template to the data.

\section{Template Code Products}\label{sec:appendix_template_code}

The RRab template for both SDSS and DES filters and the fitting algorithm presented in this work are available at \url{https://github.com/longjp/rr-templates}. The template is originally implemented in R, but can be accessed in Python via the rpy2 module as was done in this analysis. Examples of the template usage are available in both R and Python 3, though it is also compatible with Python 2.  

When using the template fitting functions, there are two options available to the user that impact the values returned by the fits. As described in the previous section, the template model includes an optional dust term. The model already includes extinction coefficients appropriate for both the DES and SDSS filter systems and can estimate the amount of extinction affecting the light curve as one of the parameters. However, if the light curves to be fit are already corrected for dust extinction prior to fitting, this parameter can be turned off to reduce the number of estimated parameters from 5 to 4 and thus improve the quality of the fits. This option is included in the code to allow the user to choose the option most appropriate for their data. 

The other option determines whether or not to use the uncertainties associated with the individual observations in the light curve when performing the fits. In this analysis, we rescaled the uncertainties and used them when fitting the template to our data. However, if one suspects that the magnitude uncertainties in the light curves are misestimated or they simply aren't available, this option can be turned off and the uncertainties will not be used. We leave this option open to the user.

The repository contains examples explaining how to fit the template to both DES and SDSS light curves using all combinations of these options. To make this code accessible to a variety of users, these examples are included as R and Python Jupyter notebooks.

\section{Data Products}\label{sec:appendix_data_products}

To enable further work with similar data, we provide all of the RRab candidate and training light curves at \url{https://des.ncsa.illinois.edu/releases/other/y3-rrl} and an extended version of Table \ref{tab:candidates}. These light curves have already had their photometric uncertainties rescaled as described in \S\ref{sec:rescale} and include both the dust-corrected and uncorrected magnitudes in each band. The light curves are indexed by their DES Y3Q2 QUICK\_OBJECT\_ID numbers, which are included as a column in the full data table.

A description of the columns in the included data table is shown in Table \ref{tab:columns} and example sample selection criteria are shown in Table \ref{tab:selectionex}. All of this information is also available in the documentation at \url{https://des.ncsa.illinois.edu/releases/other/y3-rrl}.

\begin{deluxetable}{ll}
	\tabletypesize{\footnotesize}
	\tablecaption{Description of Data Columns} 
    \label{tab:columns}
	\tablehead{\colhead{Column Name} & \colhead{Description}}
	\startdata
    QUICK\_OBJECT\_ID & DES Y3Q2 ID Number \\
    COADD\_OBJECT\_ID & DES DR1 ID Number \\
    RA & Right Ascension in degrees (J2000) \\
    DEC & Declination in degrees (J2000) \\
    p\_ab & Classifier RRab score \\
    EBV & Extinction value from Schlegel, Finkbeiner, \& Davis 1998\\
    $<g>(rizY)$ & mean $g(rizY)$ magnitude measured in light curves (not extinction corrected)\\
    nobs\_$g(rizY)$ & Number of observations in DES $g(rizY)$ in object's light curve\\
    nobs & Total number of observations in final light curve \\
    period\_0(1,2) & Period of 1st(2nd,3rd) best fitting template (days) \\
    sigma\_dp\_p & Uncertainty in $\Delta P/P$ of the best fit period \\ 
    amp\_0(1,2) & Amplitude of 1st(2nd,3rd) best fitting template (mag) \\
    sigma\_da & Uncertainty in $\Delta a$ of the best fit amplitude \\ 
    mu\_0(1,2) & Distance Modulus of 1st(2nd,3rd) best fitting template (mag) \\
    phase\_0(1,2) & Phase offset of 1st(2nd,3rd) best fitting template \\
    rss\_0(1,2) & Residual Sum of Squares (RSS) of 1st(2nd,3rd) best fit template \\
    chi2\_g(rizY) & Reduced chi squared of light curve from constant value in DES $g(rizY)$ \\
    sig\_g(rizY) & ``Significance" of light curve in DES $g(rizY)$ \\
    rss\_dof\_0(1,2) & RSS per degree of freedom of 1st(2nd,3rd) best fitting template \\
    lchi\_med & Median Log(reduced chi squared) across DES grizY \\
    rss\_lchi\_med & (RSS/dof)/ Median Log(reduced chi squared) across DES $grizY$ \\
    amp\_rss\_0(1,2) & Amplitude/(RSS/dof) for 1st(2nd,3rd) best fitting template \\
    f\_dist1\_0(1,2) & Distance of 1st(2nd,3rd) best fit Amplitude/Period from \cite{sesar2010} Oosterhoff I relation \\
    f\_dist2\_0(1,2) & Distance of 1st(2nd,3rd) best fit Amplitude/Period from \cite{sesar2010} division between the Oosterhoff I and II groups \\
    kappa\_0(1,2) & von Mises-Fisher concentration parameter of phases for 1st(2nd,3rd) best fitting template \\
    flag\_minmax & Light curve sampling flag, 0:$\ge2$ obs at max and min, 1:$<$2 obs at min, 1:$<$2 obs at max, 3:$<$2 obs at max and min \\
    GaiaDR2\_ID & Gaia DR2 source\_id for cross-matched RRab in \cite{clementini2018} \\
    PS1\_RRab & Present in \cite{sesar2017} RRab catalog? 0 = no, 1 = yes \\
    CSDR2\_ID & ID number from associated Catalina Surveys DR2 RRab catalogs \\
    ATLAS\_ID & ID from \cite{heinze2018} ATLAS RRab catalog \\
    SDSS\textunderscore ID & SDSS DR7 ID \\
    SDSS\_class & classification from S82 studies from \cite{sesar2010} or \cite{ivezic2007} \\
    Simbad\_ID & Object identifier in {\tt SIMBAD} database \citep{simbad2000} \\
    Simbad\_class & Object classification in {\tt SIMBAD} database \citep{simbad2000} \\
    OGLE\_ID & OGLE ID from LMC and SMC RRAb catalogs from \citet{ogle2016}\\
    CEMV\_ID & ID from \citet{martinezvazquez2016}\\
    GCVS5\_ID & Cross matched ID with the General Catalogue of Variable Stars v5.1 \citep{samus2017}\\
    GCVS5\_Type & Variable type for cross matched object as listed in the General Catalogue of Variable Stars v5.1 \citep{samus2017}\\    
    hpx32 & Object's Healpix (nside=32) used to subdivide the light curves \\
    comments & Comments from first author during visual validation followed by a reason code \\ 
    train & Identifies objects used to train the random forest classifier. 0 = no, 1 = yes \\
    rrab & Identified as a high confidence RRab in this study. 0 = no, 1 = yes \\
    filepath & filepath to the object's light curve \\
    \enddata
	\tablecomments{All of these features are included in the table included in the data products available at \url{https://des.ncsa.illinois.edu/releases/other/y3-rrl}.\\
    For the ``comments" column, the possible comments and their meanings are as follows:\\
    \indent \qquad confirmed -- convincing RRab\\
    \indent \qquad no\_: rejected followed by a reason code\\
    \indent \qquad maybe\_: ambiguous candidate followed by a reason code \\
    \indent \qquad missing\_image  missing coadd image in SkyViewer \\
    Reason codes:\\
    \indent \qquad galaxy -- DES image showed galaxy \\
    \indent \qquad bad\_fit -- poor template fit \\
    \indent \qquad n\_points -- few observations and/or poor phase coverage \\
    \indent \qquad crowding -- object is close to another object \\
    \indent \qquad misc -- miscellaneous reasons}
\end{deluxetable}

\begin{deluxetable}{ll}
	\tabletypesize{\footnotesize}
	\tablecaption{Example Selection Criteria}   
    \label{tab:selectionex}
	\tablehead{\colhead{To select:} & \colhead{Choose:}}
	\startdata
     Objects that were identified as RRab by the classifier & p\_ab $\ge$ 0.35 \\
     Objects found in \citet{clementini2018} Gaia DR2 RRab catalog & GaiaDR2\_ID $>$ 0 \\
     Objects found in S82 & SDSS\_class $!=$ 0 \\
     Objects found in \cite{sesar2017} Pan-STARRS catalog & PS1\_RRab $>$ 0 \\
     Objects found in Catalina Surveys RRab catalog & CSDR2\_ID $!=$ 0 \\
     Objects found in ATLAS RRab catalog & ATLAS\_ID $!=$ 0 \\
    \enddata
\end{deluxetable}

\section{Simulating DES RRab Light Curves}\label{sec:simulated}

Because DES is deeper than most overlapping surveys and much deeper than our training set in S82, we use simulated light curves to estimate the recovery rates at fainter magnitudes. We create the simulated light curves as follows: 

1) We generate light curve shapes from all 379 smoothed RRab light curve templates from \citet{sesar2010} with {\tt gatspy}'s {\tt RRLyraeGenerated} function \citep{vanderplas2015}. These generated template light curves already include the measured period and amplitude from their real observed counterpart light curves. \\

2) We shift the light curves into the DES filter system using the DES-SDSS filter transformation relations\footnote{\url{http://www.ctio.noao.edu/noao/node/5828\#transformations}}:

\begin{equation}\label{eq:filters}
  \begin{aligned}
  g_{\mathrm{DES}} &= g_{\mathrm{SDSS}} + 0.001 - 0.075(g-r)_{\mathrm{SDSS}} \nonumber  &RMS = 0.021 \  \mathrm{per\ star}\\ 
  r_{\mathrm{DES}} &= r_{\mathrm{SDSS}} - 0.009 - 0.069(g-r)_{\mathrm{SDSS}} \nonumber  &RMS = 0.021 \  \mathrm{per\ star}\\ 
  i_{\mathrm{DES}} &= i_{\mathrm{SDSS}} + 0.014 - 0.214(i-z)_{\mathrm{SDSS}} - 0.096(i-z)_{\mathrm{SDSS}}^2  \ &RMS = 0.023 \  \mathrm{per\ star}\\
  z_{\mathrm{DES}} &= z_{\mathrm{SDSS}} + 0.022 - 0.068(i-z)_{\mathrm{SDSS}}  \nonumber  &RMS = 0.025 \  \mathrm{per\ star}\\ 
  Y_{\mathrm{DES}} &= z_{\mathrm{SDSS}} + 0.045 - 0.306(i-z)_{\mathrm{SDSS}} \nonumber  &RMS = 0.030 \  \mathrm{per\ star}
  \end{aligned}
\end{equation}

3) We use the distance estimates from \citet{sesar2010} in the distance modulus equation to shift the light curves to a distance of 10 pc so that the light curves reflect the absolute magnitudes of the stars.\\

4) We then sample the photometric measurements in the light curve at phases corresponding to the real DES cadence. The cadence is randomly selected from 1808 distinct fields in the DES wide-field footprint with unique observation times. This results in a light curve that is sampled in the same manner as the light curve for a real object somewhere in the DES footprint.\\ 

5) To simulate the effects of distance on our recovery, we shift the downsampled light curve magnitudes to the apparent magnitudes they would have at a specified distance within the range that DES could detect in the single epoch images. Once the magnitudes are shifted, we remove any magnitudes that are fainter than the median magnitude depth for that filter in the DES single epoch images: $g\sim23.5,\  r\sim23.3,\  i\sim22.8,\  z\sim22.1,\  Y\sim20.7$ \citep{abbott2018}. Thus, the light curves reflect the magnitude limits of each band at fainter magnitudes. \\

6) Last, we assign a photometric uncertainty to each observation in the light curves. Following a method similar to \citet{medina2018}, we calculate the standard deviation of error-rescaled light curves in the survey region as a function of their mean magnitudes in each band as shown in Figure \ref{fig:rescale_errors}. We apply a shift in magnitude to the simulated observations using a Gaussian distribution with the appropriate standard deviation for that magnitude and band. \\

\begin{figure*}
	\includegraphics[width=1\textwidth]{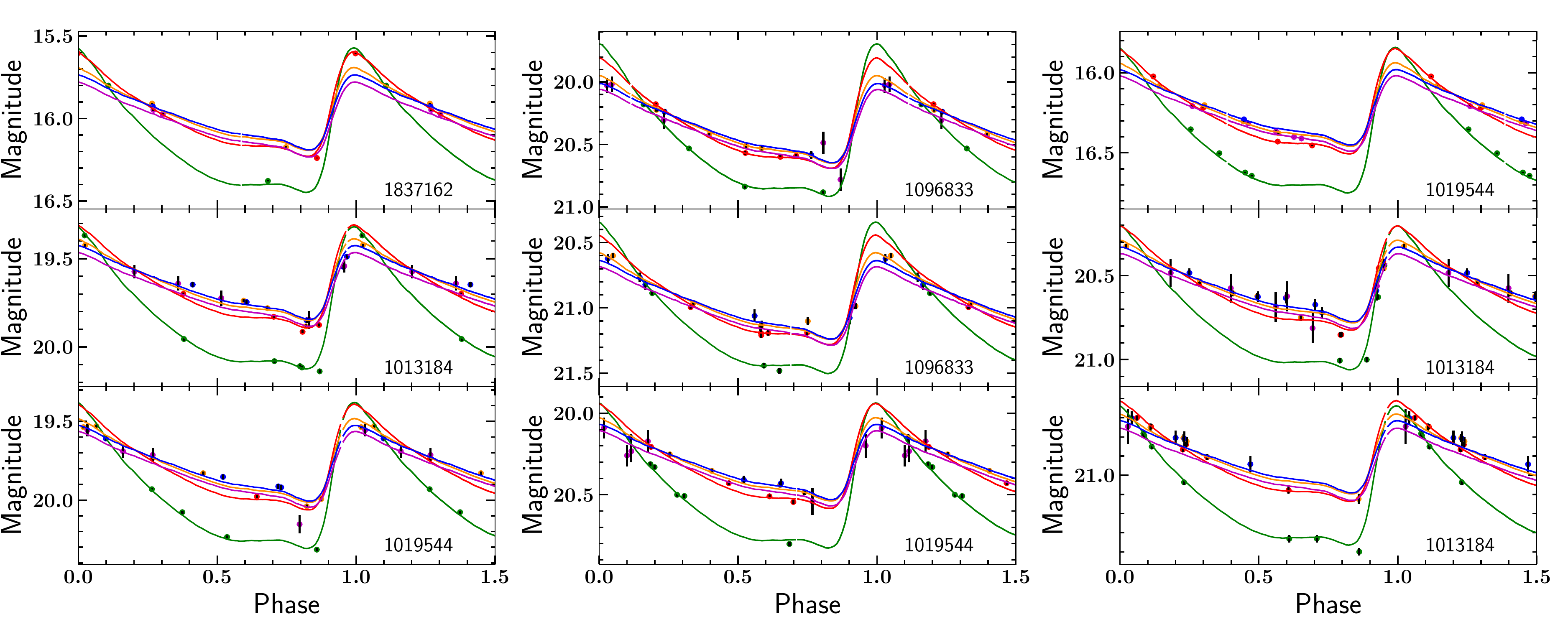}
	\caption{Examples of simulated RRL light curves labeled with the ID number of the SDSS light curve used to generate them. The observations and templates are colored by filter using the same convention as Figure \ref{fig:template}.}
    \label{fig:sim_lc}
\end{figure*}

Following this procedure, we created 5685 simulated RRab light curves. Since the photometric uncertainties were sampled from the rescaled uncertainties we applied to the real data, there was no need to rescale the errors using the method described in Section \ref{sec:rescale}. Because we did not also simulate non-variable light curves to analyze alongside the simulated RRab light curves, rescaling the errors using the same procedure would have removed real variable objects. We fit the template to all 4751 simulated light curves which passed the initial variability cuts and had at least 5 observations and used the results to determine our detection efficiency. Results of this analysis are detailed in Section \ref{sec:simresults}.


\end{document}